\documentclass[%
 reprint,
superscriptaddress,
showpacs,preprintnumbers,
 amsmath,amssymb,
 aps,
pre,
]{revtex4-1}

\usepackage{dcolumn}
\usepackage{bm}

\usepackage{graphicx,capt-of}
\usepackage{color}
\usepackage{bm}
\usepackage{subfigure}
\usepackage{caption}
\usepackage{tabularx,ragged2e,booktabs,caption}
\usepackage{grffile}
\usepackage{mathrsfs}
\usepackage{multirow}
\usepackage{amsmath}
\usepackage{mathtools}

\def\beq{\begin{equation}}
\def\eeq{\end{equation}}
\def\ba{\begin{eqnarray}}
\def\ea{\end{eqnarray}}

\begin{document}


\title{Relaxation dynamics of the three-dimensional Coulomb Glass model}

\author{Preeti Bhandari}
\affiliation{Department of Physical Sciences, Indian Institute of Science Education and Research (IISER) Mohali, Sector 81, S.A.S. Nagar, Manauli P. O. 140306, India}
\affiliation{Department of Physics, Ben Gurion University of the Negev, Beer Sheva 84105, Israel}
\author{Vikas Malik}%
 \email{vikasm76@gmail.com}
\affiliation{Department of Physics and Material Science, Jaypee Institute of Information Technology, Uttar Pradesh 201309, India.}

\author{Deepak Kumar}
\email{deceased}
\affiliation{
 School of Physical Sciences, Jawaharlal Nehru University, New Delhi -- 110067, India.}
\author{Moshe Schechter}
\affiliation{Department of Physics, Ben Gurion University of the Negev, Beer Sheva 84105, Israel} 

\date{\today}

\begin{abstract}
 In this paper, we analyze the dynamics of the Coulomb Glass lattice model in three dimensions near a local equilibrium state by using mean-field approximations. We specifically focus on understanding the role of localization length ($\xi$) and the temperature ($T$) in the regime where the system is not far from equilibrium. We use the eigenvalue distribution of the dynamical matrix to characterize relaxation laws as a function of localization length at low temperatures. The variation of the minimum eigenvalue of the dynamical matrix with temperature and localization length is discussed numerically and analytically. Our results demonstrate the dominant role played by the localization length on the relaxation laws. For very small localization lengths we find a crossover from exponential relaxation at long times to a logarithmic decay at intermediate times. No logarithmic decay at the intermediate times is observed for large localization lengths.
\end{abstract}

\pacs{71.23.Cq, 73.50.-h, 72.20.Ee}
\maketitle


\section{\label{sec:level1}Introduction}

The term Coulomb Glass (CG) refers to that category of disordered insulators that have a sufficiently high disorder, which leads to localized electronic states coupled with the Coulomb interactions. The presence of a glassy phase in this model has been predicted theoretically by several authors \cite{jpt82,mbld82,mm82,m84,ec94}. In dimensionless units, the Hamiltonian for CG lattice model is defined \cite{ab75} as

\begin{equation}
\label{Hamiltonian}
\mathcal{H} \{n_{i}\} = \sum_{i=1}^{N} \epsilon_{i} n_{i} +  \frac{1}{2} \sum_{i \neq j} \frac{e^{2}}{\kappa |\vec{r_{i}} - \vec{r_{j}}|} (n_{i} - 1/2) (n_{j} - 1/2)
\end{equation}  
Where, $\epsilon_{i}$'s are the on-site random field energy and the occupation number $n_{i} \in \{0,1\}$. The electrons at site $i$ and $j$ interact via unscreened Coulomb interaction $e^{2}/(\kappa \ r_{ij})$ where $\kappa$ is the dielectric constant.

 Much work has been done to find the ground state of the CG model at high disorder. Using mean field approach \cite{mbld82}, Monte Carlo simulation \cite{amb92,pvs17,pv17} and other optimization \cite{sabb79,avjmy08} approaches it has been found that there exist many metastable states (pseudo ground states) at low temperatures. This metastability is responsible for glassy behavior. The density of states (DOS) found in all these approaches shows a soft gap $g(E) \sim E^{\delta}$ around the Fermi level \cite{ab75,ba84,mma13,m70,g71,jpt84,jm95,vjp00}. The value of $\delta$ is very near to the theoretical prediction of  $d-1$ ( $d$ is the dimensionality of the system) given by Efros and Shklovskii \cite{ab75}.  Recently M\"{u}ller and Ioffe have established a connection between the presence of a glassy phase and the appearance of a soft gap in three dimensional CG model using locator approximation \cite{mi04}. The formation of a gap in the DOS affects the conductivity ($\sigma$) quiet significantly. One can see that the conductivity changes from the Mott's law \cite{m68,m69} of  $ln \ \sigma \sim (T_{M}/T)^{1/4}$ to the Efros-Shklovskii's law $ln \ \sigma \sim \ (T_{ES}/T)^{1/2}$ law \cite{ab75} at low temperatures. 

  The existence of glass transition in three dimensional CG has been controversial and is a matter of active research \cite{mm09,bhtag09,am10,ajmh19}. Although, some mean-field analysis, supported by recent numerical analysis \cite{ajmh19}, do suggest the presence of a stable glassy phase \cite{av99,sv05,mi04,ms07,am82}.  Non-equilibrium dynamics of structural and spin glasses has been studied using scaling properties of non-stationary correlation and response functions \cite{l78,e97,jlj97,l02,lj93,lj94}. Various numerical simulations claim that the CG model exhibits glassy behavior i.e. slow relaxation \cite{azm000,add05,mj09,ayy09,ayy10,jy12}, aging \cite{ayy09,azm00} and memory effects \cite{vz04,azm02}. Many experimental techniques are used to study relaxation in the CG model \cite{mzm93,gdcanya97,gcdnaa98,zm97,azm98,azm00,azm02,vz04,z17,z18,z19,vz07,jtcvll20} . The basic idea is to introduce a perturbation in the material to push the system out of equilibrium. This leads to an increase in the conductance, whose decay with time is then measured. It has been observed that many materials, amorphous as well as crystalline, show a logarithmic temporal decay in conductance. 

The study of slow relaxation can be categorized broadly into two types of models: $(a)$ A quasi-particle model, which was proposed by Pollak and Ovadyahu \cite{mma13,mz03}. They considered multi-particle transitions and showed that the decrease in energy with time is related to $\gamma_{m}$ which is the minimal value of the transition rates $\gamma = \tau_{0}^{-1} exp[-\frac{r}{\xi}-\frac{E}{kT}]$ where $r$ and $E$ are the collective hopping distance and energy respectively and $\xi$ is the localization length. Assuming that the change in conductance ($\Delta G(t)$) and energy are related to each other linearly one gets a logarithmic decay in conductance
\begin{equation}
\label{mzeq}
\Delta G(t) \propto -ln(\gamma_{m} \ t)
\end{equation}
$(b)$ Second is the local mean-field model, suggested by Amir {\it et al} \cite{ayy09,ayy10,ayy08}. The dynamics of quite a few systems near local stable minima can be described by the matrix equation
\begin{equation}
\label{amireq}
\frac{d\delta n}{dt} = -A \delta n
\end{equation} 
where $\delta n_{i}= n_{i}-f_{i}$ is the fluctuation of the occupation number ($n_{i}$) from its value $f_{i}$ at the local minima. Amir {\it et al} have shown \cite{ayy08} that under mean-field approximations and single-particle transitions dynamics, the CG model obeys Eq.(\ref{amireq}). The regime of low temperatures and small localization lengths is considered, and the distribution $P(\lambda) \sim \frac{1}{\lambda}$ is found for the small relaxation rates. This leads to a logarithmic decay of fluctuations in occupation numbers $\delta n(t)$. Assuming that the relaxation of excess conductance $\Delta G(t)$ is linear in $\delta n(t)$, one recovers the logarithmic decay for conductance as given in Eq.(\ref{mzeq}). In this approach, the system always remains near the local minima, and thus the transition between different metastable states (multi-particle transitions) is completely neglected.

 Our goal here is to study the relaxation effects in the Coulomb Glass lattice model near a local equilibrium state by using mean-field approximations. We follow the approach of Amir $\it{et}$ $\it{al}$ \cite{ayy08}, albeit for a lattice CG model. Within the approach of Amir $\it{et}$ $\it{al}$ there is disorder in site energies as well as in the position of the sites. In their approach \cite{ayy08}, and small localization lengths studied, the slow dynamics are mainly due to isolated localized states that have a long life-time. However, in the lattice model discussed here, disorder comes only via site energies and so the question of isolated states does not come into the picture. Instead, we find that for small localization lengths, $\xi \ll 1$, the states near the Fermi level are very stable and any fluctuations in them relax very slowly. The main reason for this slow decay is that the states near the Fermi level are isolated energetically due to the hard gap in the DOS inflicted on their near neighbor sites. For all localization lengths and temperatures, the system always obeys the exponential relaxation ($\delta n(t) \sim exp-(t/\tau_{max})$), at times longer than $\tau_{max}$. The maximum relaxation time ($\tau_{max}$) is inversely proportional to the smallest eigenvalue ($\lambda_{min}$) of the dynamical matrix $A$. Our study shows that $\lambda_{min}$ depends upon the localization length as well as temperature. 
 
 We further find that logarithmic time dependence of the relaxation of $\delta n(t)$ at intermediate times is present only for small localization lengths, $\xi \ll 1$, where relaxation is mainly due to jumps to nearest neighbor sites.  
   
 The paper is organized as follows. In Sec. \ref{sec:level2}, we have provided an overview of our derivation of the linear dynamical matrix. In Sec. \ref{sec:level3}, we present a detailed discussion of our mean-field results obtained numerically and analytically. And finally in Sec. \ref{sec:level4}, we provide the conclusions of our work.
  
 \section{\label{sec:level2}Dynamics}
 
 The most general non-conserved dynamics for the total probability distribution of the spins was developed by Glauber \cite{car94}. This was extended to conserved dynamics by Kawasaki who incorporated the constraint of fixed magnetization. The Kawasaki formulation \cite{rp96,aa93} applies to CG as the electron number is conserved - which is equivalent to fixed magnetization. Here we deal with the probability distribution of $P(n_{1} \ldots n_{N};t)$, which involves the occupation of all sites in the system. The Kawasaki dynamics holds for the interacting system as well as for multi-particle dynamics. Since this approach is general, it can be taken beyond mean-field theory. 
 
 The time evolution of a system can be described using a generalized master equation \cite{pw09}
\ba
\label{eq1}
\dfrac{d}{dt} P (\lbrace n_{i}\rbrace, t) = - \sum_{i \neq j}  \hspace*{2mm} W_{i \rightarrow j} \hspace*{2mm} P(\lbrace n_{i}\rbrace,t) \nonumber \\
+ \sum_{j \neq i}  \hspace*{2mm} W_{j \rightarrow i} \hspace*{2mm} P(\lbrace n_{j}\rbrace,t)
\ea
where $W_{i \rightarrow j}$ denotes the transition rates from state $i$ to $j$ and $P (\lbrace n_{i}\rbrace, t)$ is the probability of finding the system in state $i$ at time $t$. The transition rates can be single or multi-electron transfer. Since we are interested in Kawasaki dynamics, only transitions that conserve the particle (electron) number will be considered. Using single-particle transitions, the Kawasaki dynamics equation can be rewritten as 
\begin{equation}
\label{eq2}
\begin{aligned}
\dfrac{d}{dt}P(n_{1} \ldots n_{\nu};t)={} &  -\sum_{i \neq j} \omega_{i \rightarrow j} \hspace*{2mm} n_{i}(1-n_{j}) \hspace*{2mm}\times \\
 & P(\ldots,n_{i} \ldots n_{j},\ldots;t) \hspace*{2mm}+  \\
&   \sum_{i \neq j} \omega_{j \rightarrow i} \hspace*{2mm} \bar{n}_{j}(1-\bar{n}_{i})\hspace*{2mm}\times \\
& P(...,\bar{n}_{i}...,\bar{n}_{j},...;t)
\end{aligned}
\end{equation}
where $\omega_{i \rightarrow j}$ is the transition probability from site $i$ to $j$ and $\bar{n}_{i}=1-n_{i}$. Now we impose the condition of "detailed balance", so that the evolution is towards thermal equilibrium. In thermal equilibrium,
\begin{multline}
\label{eq3}
\omega_{i \rightarrow j} \hspace*{2mm} n_{i}(1-n_{j})  \hspace*{2mm} P^{eq}(\ldots,n_{i} \ldots n_{j},\ldots) = \\
  \omega_{j \rightarrow i}  \hspace*{2mm} \bar{n}_{j}(1-\bar{n}_{i})  \hspace*{2mm} P^{eq}(\ldots,\bar{n}_{i} \ldots,\bar{n}_{j},\ldots).
\end{multline}
\begin{equation}
\label{eq4}
\frac{\omega_{i \rightarrow j}}{\omega_{j \rightarrow i}} = \frac{exp [-\beta E(\ldots,\bar{n}_{i} \ldots,\bar{n}_{j},\ldots)]}{exp [-\beta E(\ldots,n_{i} \ldots,n_{j},\ldots)]} \, .
\end{equation}
The energy required to transfer an electron from $i$ to $j$ is
\ba
\label{eq5}
\Delta E_{ji} &=& E(\ldots,\bar{n}_{i} \ldots,\bar{n}_{j},\ldots)) - E(\ldots,n_{i} \ldots,n_{j},\ldots), \nonumber \\
&=& \epsilon_{j} -\epsilon_{i} + \sum_{m \neq i} K_{jm} n_{m} - \sum_{m \neq j} K_{im} n_{m},  \nonumber \\
&=& \widetilde{E^{i}_{j}} - \widetilde{E^{j}_{i}}. \quad 
\ea
where
\ba
\label{eq6}
\widetilde{E^{i}_{j}} &=& \epsilon_{j} + \sum_{m \neq i} K_{jm} n_{m} \, , \nonumber \\
&=&   E_{j} - K_{ji} n_{i} \quad 
\ea 
and $E_{j}$ is the Hartree energy: $E_{j} = \epsilon_{j} + \sum_{m} K_{jm} n_{m}$ and $K_{jm}=\frac{1}{r_{jm}}$ is the Coulomb interaction term. We can then rewrite Eq.(\ref{eq5}) as
\ba
\label{eq7}
\Delta E_{ji} &=& E_{j} - E_{i} - K_{ij} (n_{i} - n_{j}), \nonumber \\
&=& E_{j} - E_{i} - K_{ij}  \quad 
\ea
hence we get $\frac{\omega_{i \rightarrow j}}{\omega_{j \rightarrow i}} = e^{-\beta \Delta E_{ji}}$. So we choose our transition probability as
\begin{equation}
\label{eq8}
\omega_{i \rightarrow j} = \frac{\gamma_{ij}}{2 \tau} \frac{1}{e^{\beta \Delta E_{ji}}+ 1}
\end{equation} 
where $\tau$ is a hopping time scale.
With this choice, master equation takes the form
\ba
\label{eq9}
\dfrac{d}{dt}P(\lbrace n_{l}\rbrace;t)=  -\sum_{i \neq j} \frac{\gamma(r_{ij})}{2 \tau} n_{i}(1-n_{j}) \nonumber \\ 
\Big[ f(\Delta E_{ji}) P(\ldots,n_{i},\ldots,n_{j},\ldots;t) - \nonumber \\
 f(\Delta E_{ij}) P(\ldots,\bar{n}_{i},\ldots,\bar{n}_{j},...;t) \Big]
\ea
Here $f(E) = \frac{1}{exp(\beta E) + 1}$ is the Fermi-Dirac distribution, $\gamma_{ij} = \gamma(r_{ij})$ is a factor independent of temperature, but depends on the distance between sites $i$ and $j$. For hopping electrons $\gamma_{ij} = \gamma_{0} e^{-r_{ij}/\xi}$, where $\gamma_{0}$ is a constant.

From this, one can derive an equation for time-dependent averages or moments. To connect to the one-particle master equation, we consider
\begin{equation}
\label{eq10}
N_{i}(t) = \sum_{\lbrace n_{l}\rbrace} n_{i} \hspace*{1mm} P(n_{1}\ldots n_{N};t)
\end{equation} 
whose time derivative gives  
\begin{equation}
\label{eq11}
\begin{aligned}
\dfrac{d}{dt}N_{i}(t)={} &  -\frac{1}{2\tau} \sum_{j \neq k} \gamma (r_{j k}) \hspace*{2mm} \sum_{\lbrace n_{l} \rbrace} n_{i} n_{j} (1-n_{k}) \hspace*{2mm} \\
& [f(\Delta E_{kj}) \hspace*{2mm}P(n_{1}\ldots n_{v};t)  \\
&  - f(\Delta E_{jk}) \hspace*{2mm}P(n_{1}\ldots \bar{n}_{j}\ldots \bar{n}_{k},\ldots;t)]     
\end{aligned}
\end{equation} 
Again, if $i \neq j$ or $i \neq k$, a change of summation variables $j \rightleftharpoons k$ makes the two terms cancel. The only surviving term comes from $i = j$, Eq.(\ref{eq11}) can now be written as
\begin{equation}
\label{eq12}
\begin{aligned}
\dfrac{d}{dt}N_{i}(t)={} &  -\frac{1}{2\tau} \sum_{k \neq i} \gamma (r_{ik}) \hspace*{2mm} \sum_{\lbrace n_{l} \rbrace} n_{i} (1-n_{k}) \hspace*{2mm} \\ & \Big[f(\Delta E_{ki}) \hspace*{2mm}P(\ldots,n_{i}\ldots n_{k},\ldots;t)  \\
&  - f(\Delta E_{ik}) \hspace*{2mm}P(\ldots,\bar{n}_{i}\ldots \bar{n}_{k},\ldots;t)\Big]      \\
 ={} &  -\frac{1}{2\tau} \sum_{k \neq i} \gamma (r_{i k}) \hspace*{2mm} [  \left\langle n_{i} (1-n_{k}) \hspace*{2mm}  f(\Delta E_{ki})\right\rangle_{t}  \\
 & - \left\langle (1-n_{i})n_{k}  \hspace*{2mm} f(\Delta E_{ik})\right\rangle_{t}]   \, ,   
\end{aligned}
\end{equation} 
where $\langle ...\rangle_{t}$ denotes average at time $t$. The Eq.(\ref{eq12}) is an exact equation. To get a closed set of equations, one needs to apply mean-field approximation to Eq.(\ref{eq12}).

\begin{figure}
\centering
\subfigure{\includegraphics[width=5cm,angle=-90]{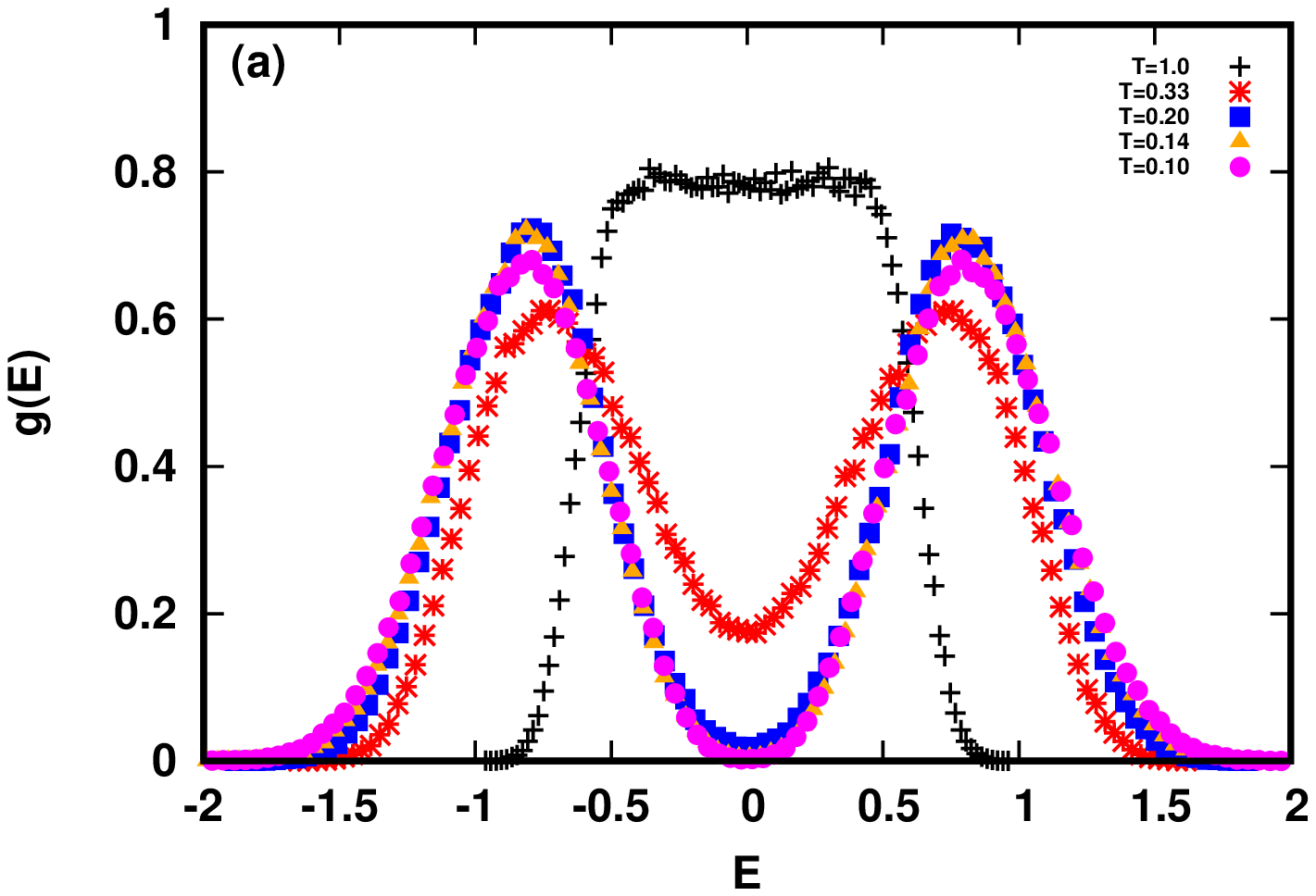}}
\subfigure{\includegraphics[width=5cm,angle=-90]{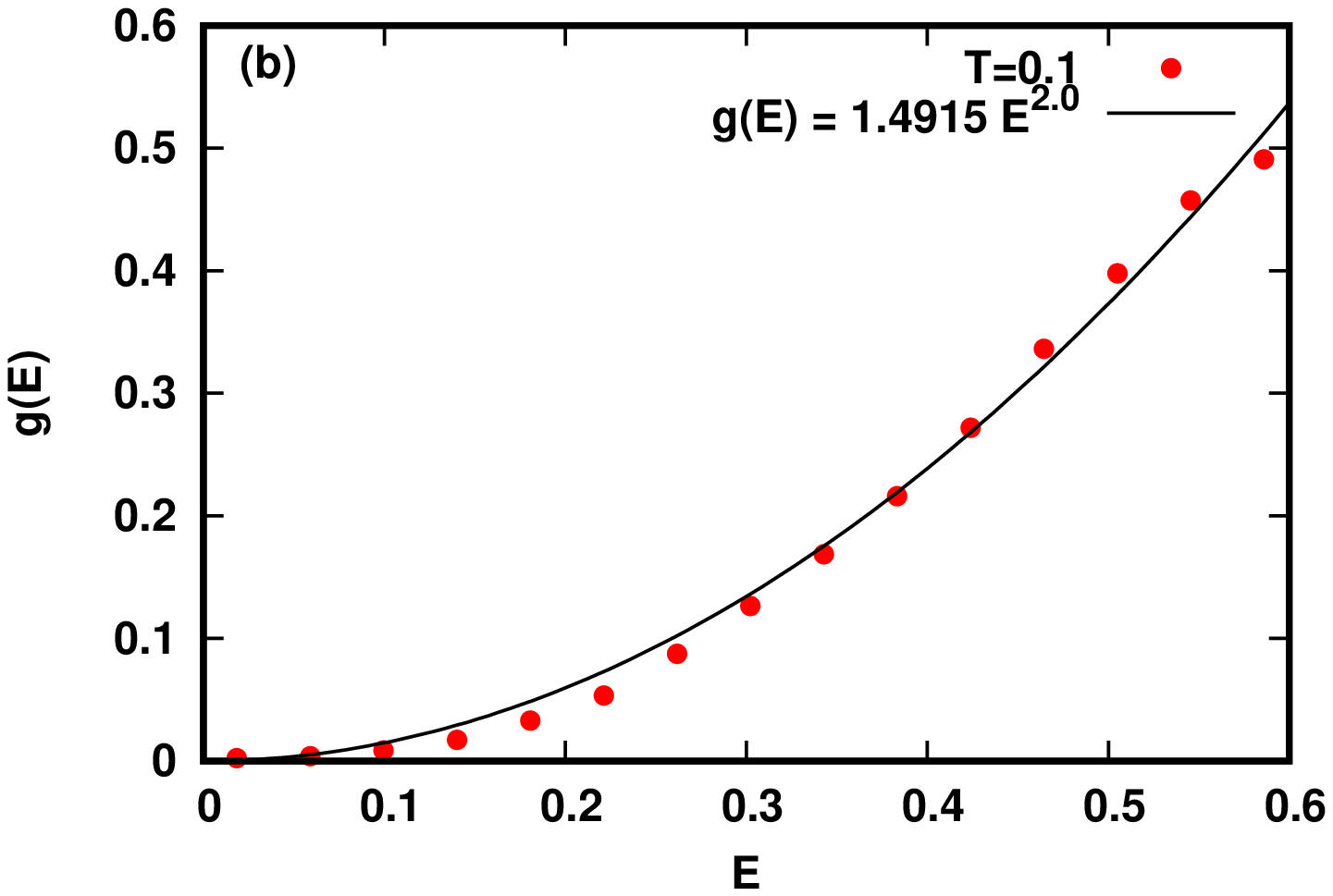}}
\caption{\label{fig:1} (Colour online) (a) Histogram of the Hartree energies $E$ (obtained using Eq. (\ref{eq16}) where $\epsilon_{i}$ were chosen from a box distribution of width $\pm \frac{W}{2}$ with $W = 1$) at different temperatures, for $L = 16$ and $\mu = 0$. (b) Zoom in to low energies of the Hartree energies for $T = 0.1$. Solid line is the best-fit $g(E) \propto E^{2.0}$.}
\end{figure}

\subsection*{\underline{Mean-Field Approximation}}
 
The mean-field approximation consists of making the assumption
\begin{eqnarray}
\label{eq13}
\left\langle f(n_{1}...n_{N};t) \right\rangle =  f(N_{1}(t)...N_{N}(t))
\end{eqnarray} 
With this assumption we get
\begin{equation}
\label{eq14}
\begin{aligned}
\dfrac{d}{dt}N_{i}(t)={} &  -\frac{1}{2\tau} \sum_{k \neq i} \gamma (r_{i k}) \hspace*{2mm} [N_{i} (1 - N_{k}) \hspace*{2mm}  f_{FD}({E}_{k}-{E}_{i})  \\
& -  N_{k}(1 - N_{i}) \hspace*{2mm}   f_{FD}({E}_{i}-{E}_{k})]     
\end{aligned}
\end{equation}
where ${E}_{i}$ and ${E}_{k}$ are the Hartree energies at site $i$ and $k$ respectively. $f_{FD}(E)=1/(exp[\beta E]+1)$ is the Fermi Dirac distribution. Now let us linearize this equation about an equilibrium solution:
\begin{equation}
\label{eq15}
N_{i}(t) = f_{i} + \delta N_{i}
\end{equation}
\begin{equation}
\label{eq16}
E^{e}_{i} = \epsilon_{i} + \sum_{l} K_{il} f_{l}
\end{equation}
where $f_{i} = \frac{1}{exp(\beta E^{e}_{i})+1}$. Putting Eq.(16) and Eq.(17) into Eq.(15) one gets: 
\begin{equation}
\label{eq17}
\begin{aligned}
\dfrac{d}{dt} \delta N_{i}={} & -\frac{1}{2 \tau} \sum_{k \neq i}  \gamma_{i k} \Bigg[(f_{i} + \delta N_{i}) (1-f_{k}-\delta N_{k}) \\
& f_{FD}(E^{e}_{k} - E^{e}_{i} + \sum_{l} (K_{kl} - K_{il}) \delta N_{l}) \\
& - (f_{k} + \delta N_{k}) (1-f_{i}-\delta N_{i}) \\
&  f_{FD}(E^{e}_{i} - E^{e}_{k} + \sum_{l} (K_{il} - K_{kl}) \delta N_{l}) \Bigg]  
\end{aligned}
\end{equation}
And the final linear equation using the detailed balance is:
\begin{equation}
\label{eq18}
\begin{aligned}
\dfrac{d}{dt} \delta N_{i}={} &  \sum_{k \neq i} \Bigg[ \frac{\delta N_{i}}{f_{i}(1-f_{i})} \hspace*{2mm} \Gamma_{i k} \hspace*{2mm} - \hspace*{2mm}  \frac{\delta N_{k}}{f_{k}(1-f_{k})} \hspace*{2mm} \Gamma_{k i}  \\
&  + \hspace*{2mm} \frac{1}{T} \sum_{k \neq l,i} \Gamma_{i k} \hspace*{2mm} (K_{k l} - K_{i l}) \hspace*{2mm} \delta N_{l} \Bigg ] \\
={} & \sum_{l} A_{i l} \hspace*{2mm} \delta N_{l}
\end{aligned}
\end{equation} 
where we define
\begin{subequations}
	\label{19}
	\begin{equation}
	\Gamma_{ik} = \frac{1}{2 \tau} \gamma(r_{i k}) \hspace*{2mm} f_{i} (1-f_{k}) \hspace*{2mm} f_{FD}(E^{e}_{k} - E^{e}_{i})
	\end{equation}  
	\begin{equation}
	\Gamma_{ k i} = \frac{1}{2 \tau} \gamma(r_{k i}) \hspace*{2mm} f_{k} (1-f_{i}) \hspace*{2mm} f_{FD}(E^{e}_{i} - E^{e}_{k})
	\end{equation}
    \begin{equation}
	A_{i i}  = -\sum_{k \neq i}\, \frac{\Gamma_{i k}}{f_{i}(1-f_{i})} 
	\end{equation}
	\begin{equation}
	A_{i l} = \frac{\Gamma_{l i}}{f_{l}(1-f_{l})} + \frac{1}{T} \sum_{k(\neq l\neq i)}\, \Gamma_{i k} \hspace*{2mm} (K_{k l}-K_{i l})
	\end{equation}
\end{subequations}  
It is easy to verify that $\Gamma_{ik} =\Gamma_{ki}$. 
Thus the final linear equation has the same form as the one used by Amir $\it{et}$ $\it{al}$\cite{ayy08}. Here $A$ is the linear dynamical matrix governing the dynamics of the system near equilibrium and $\Gamma_{ik}$ are the equilibrium transition rates. The transition rates as defined in Eq.\ref{19}(a) and Eq.\ref{19}(b) can be written as
\begin{equation}
\label{Tik}
\Gamma_{i k} = \gamma_{0} \, exp\bigg(\frac{-r_{ij}}{\xi} \bigg) \, exp\bigg(\frac{-1}{2T} [|E_{i}|+|E_{j}|+|E_{i}-E_{j}|]\bigg) \, ,
\end{equation} 
when the energies $|E_{i}|$, $|E_{j}|$ and $|E_{i}-E_{j}|$ are greater than $T$.

\begin{figure}
\centering
\subfigure{\includegraphics[width=5cm,angle=-90]{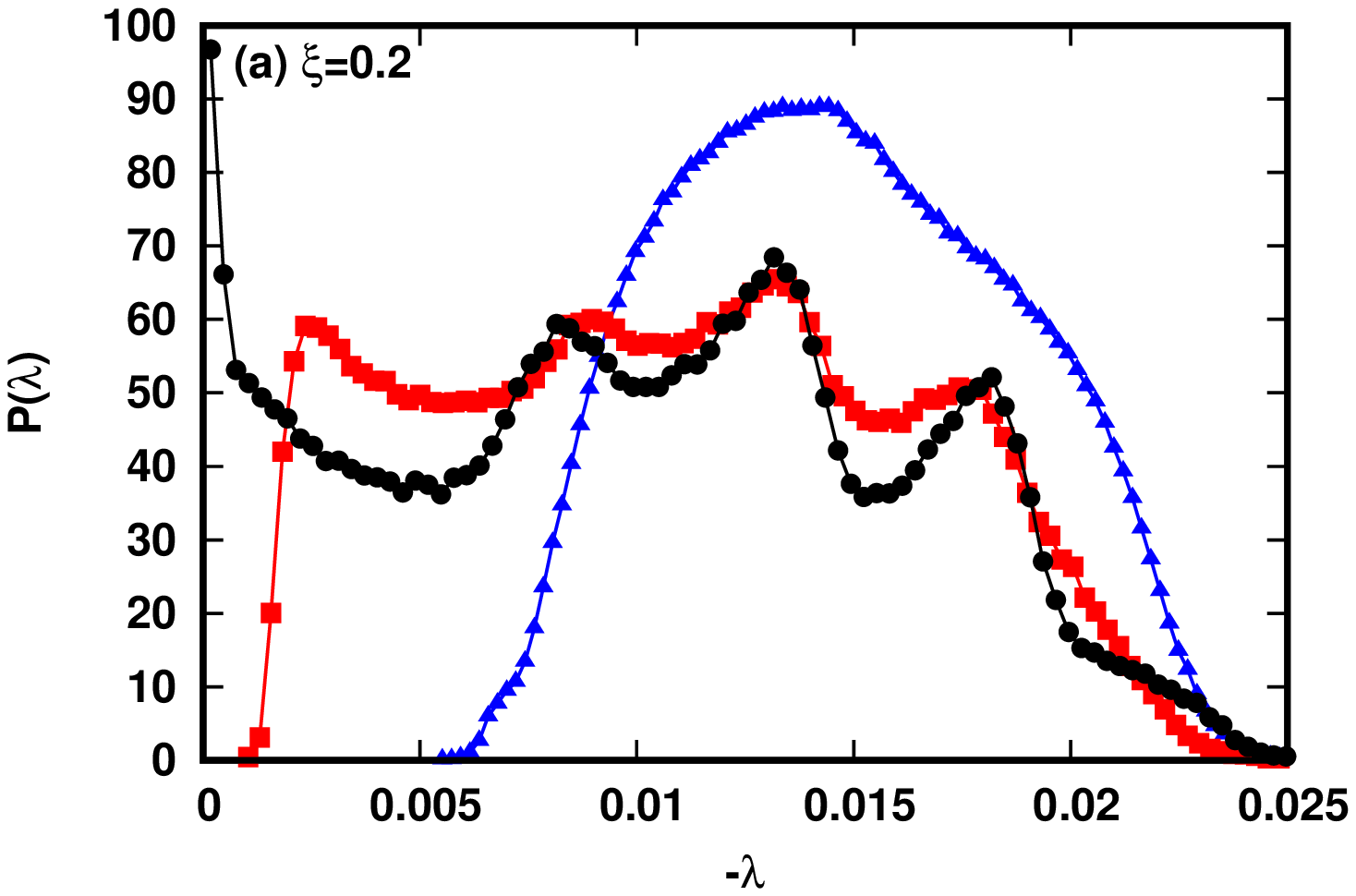}}
\subfigure{\includegraphics[width=5cm,angle=-90]{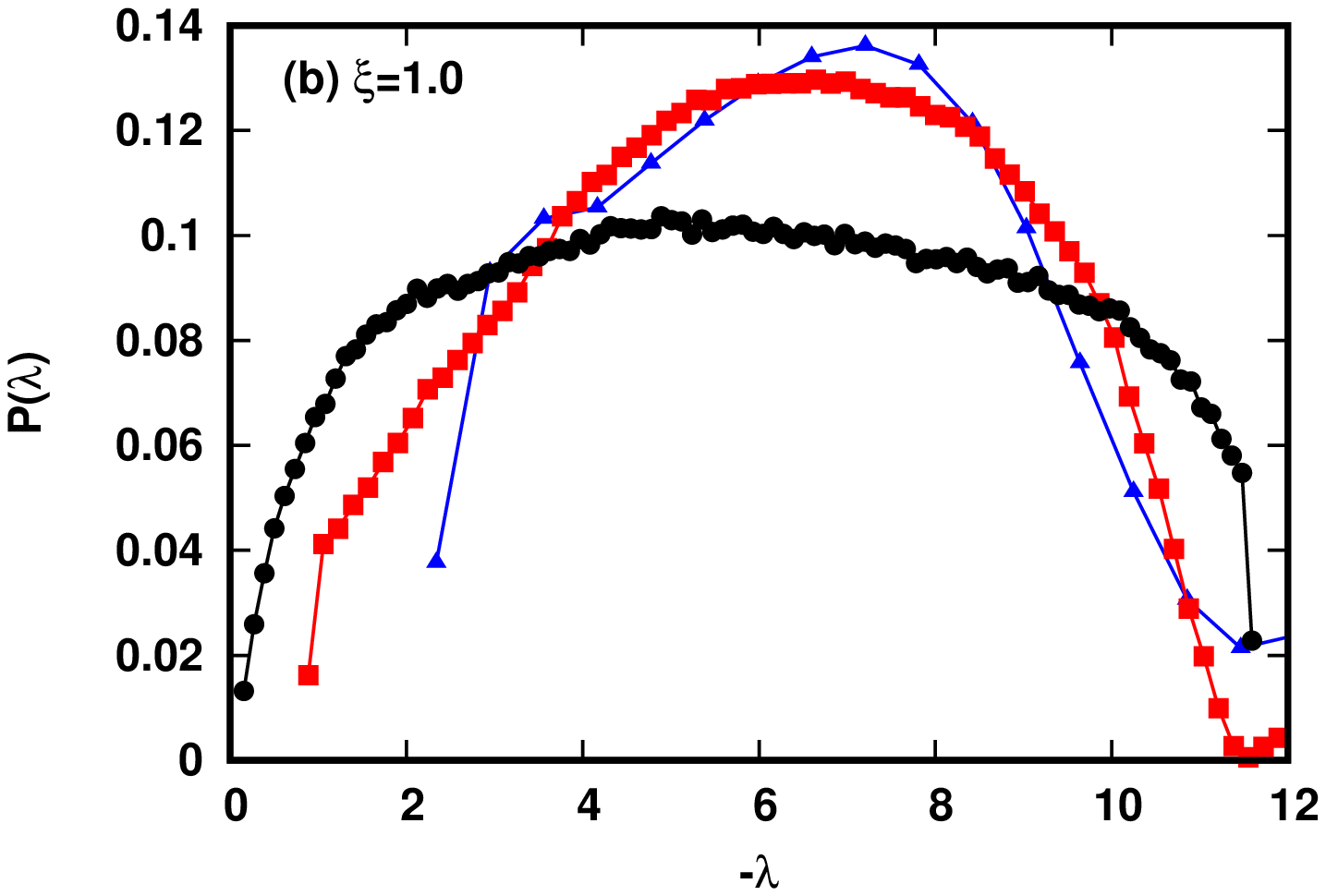}}
\caption{\label{fig:2} Distribution of the eigenvalues of the dynamical matrix $A$ obtained by solving Eq.(\ref{19}), at different temperature ($T = 0.33$ in blue, $T = 0.20$ in red and $T = 0.10$ in black) for $\xi = 0.2$ (a) and for $\xi = 1.0$ (b).}
\end{figure}
   
 \section{\label{sec:level3}Results and discussions}
  In this paper we study a three-dimensional cubic lattice of localized states which have random energies and interact through Coulomb interactions. We model this system by a Hamiltonian as defined in Eq.(\ref{Hamiltonian}). We take the number of electrons to be half of the total number of sites in the lattice. All energies are noted in units of $e^{2}/\kappa a$ where $a$ is the lattice constant.  
 \subsection{\underline{Coulomb Gap}}  
 \textit{The method-} To calculate the Hartree energy ($E_{i}$) given in Eq. (\ref{eq16}), we have first calculated the magnetization, which, approximated within the mean-field theory is defined as 
 \begin{equation}
 \label{mag_mft}
 m_{i} = tanh \ \beta \ \Bigg(E_{i} + \sum_{k} \frac{m_{k}}{r_{ik}}\Bigg)
 \end{equation}
 The above equation was solved self-consistently and the final $m_{i}$'s were then used to calculate $E_{i}$'s using $f_{i} = (m_{i}+1)/2$. We have annealed our data from $T = 1$ to $T = 0.1$, and the on-site energy $\epsilon_{i}$ was chosen randomly from a box-distribution of width $\pm W/2$ where $W = 1$ and $\beta = 1/T$.
 
 It is well established now that in the CG model, a soft gap, also called the Coulomb gap, is observed in single-particle DOS at low temperatures. The gap gets filled as the temperature increases. In this paper, the temperatures where the soft gap is well established are referred to as low temperatures (i.e. $T = 0.1-0.2$). Efros and Shklovskii\cite{ab75} have further argued that at zero temperature, the DOS follows the relation $g(E) \approx E^{d-1}$ in d-dimensional CG model. In Fig.\ref{fig:1}(a), one can see formation of a soft gap in the DOS at temperature lower than 0.33. We further found that at $T=0.1$, the DOS can be well fitted by the relation $g(E) \propto  E^{2}$ (see Fig.\ref{fig:1}(b)) as suggested by Efros and Shklovskii.   
 
 \begin{figure}
 	\includegraphics[scale=0.3]{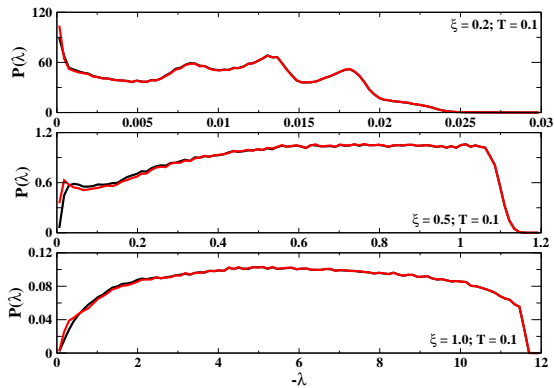}
 	\caption{\label{fig:3} (Colour online) Comparison of the eigenvalue distribution of the full A-matrix (in black) with the ones obtained after neglecting the second term in Eq.\ref{19}(b) (in red) for $\xi=0.2,0.5,1.0$, and $T=0.1$.}
 \end{figure}  

 \subsection{\underline{Linear Dynamical Matrix}}
  
 In Fig.\ref{fig:2}, we show the distribution of the eigenvalues of a linear dynamical matrix ($A-matrix$) at different temperatures and localization lengths. The eigenvalues ($\lambda$) here determine the rate of decay in the system. With the decrease in temperature, the shifting of $\lambda$ towards zero indicates a slowing down of relaxation.
 
\begin{figure}
\centering
\subfigure{\includegraphics[width=5cm,angle=-90]{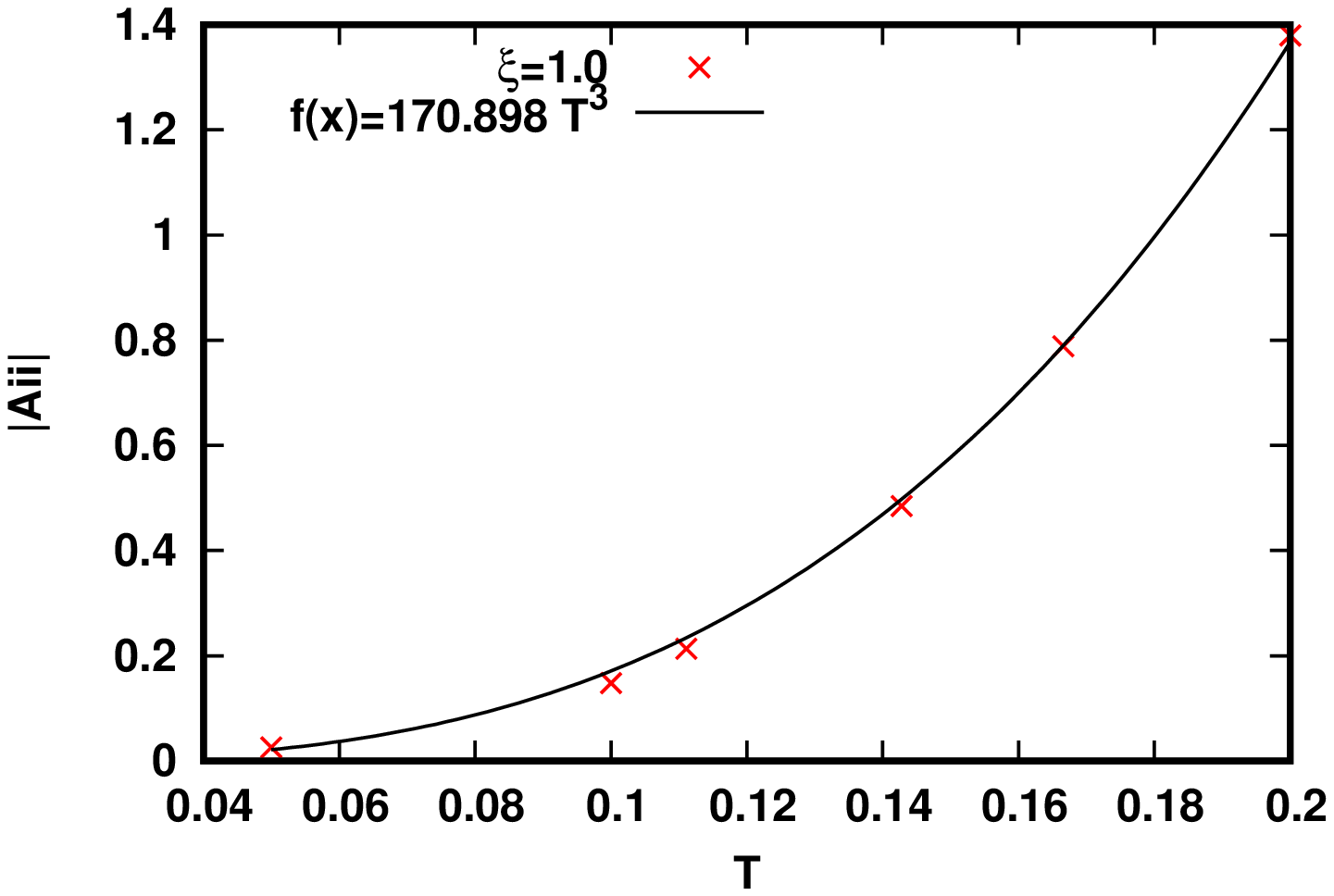}}
\subfigure{\includegraphics[width=5cm,angle=-90]{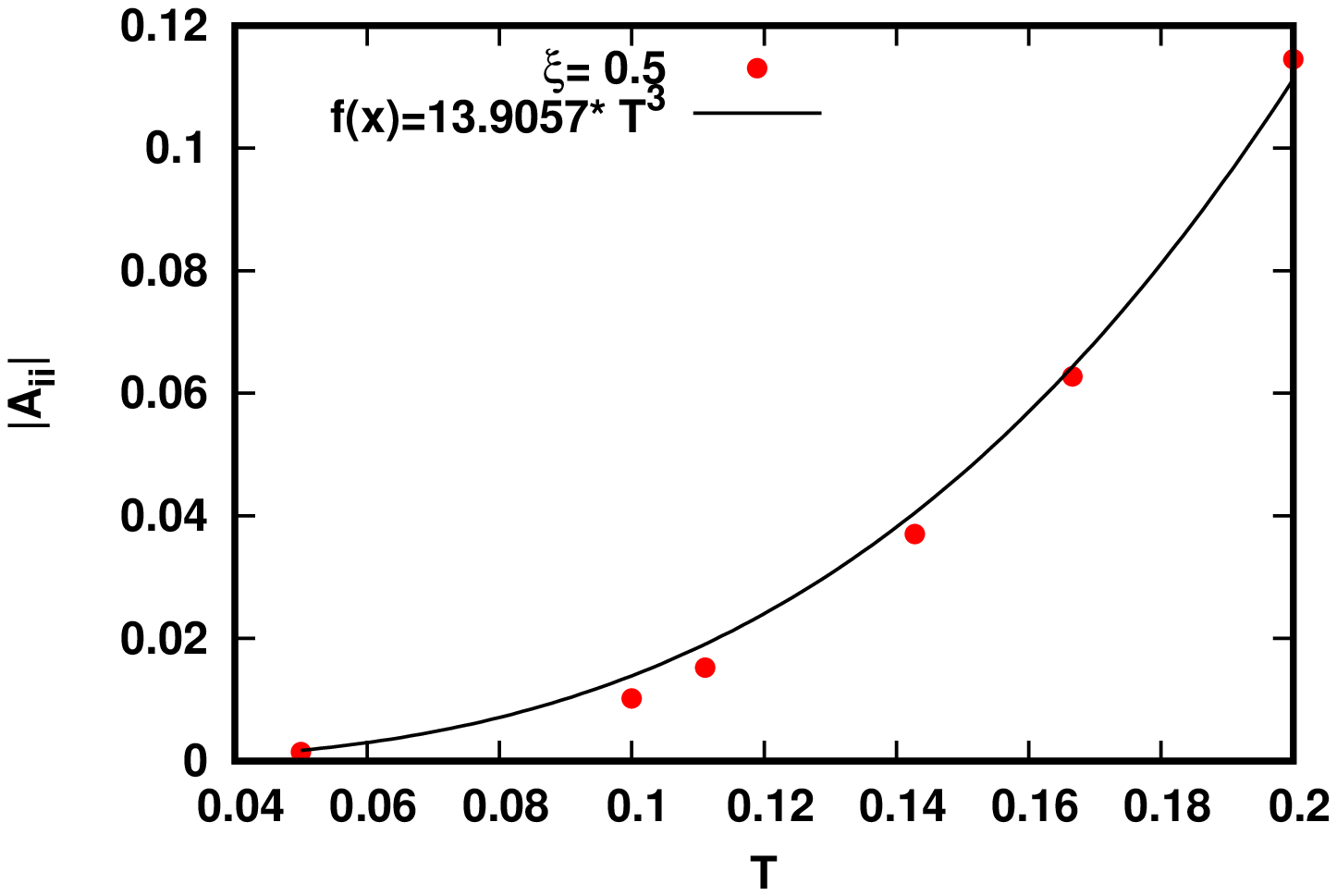}}
\caption{\label{fig:4} (Colour online) Lowest value of the diagonal part of $A-matrix$ at $\xi = 0.5,1.0$ calculated using Eq.(\ref{19}(c)), which is averaged over 100 configurations are plotted against temperature. The solid line shows that the data fits well with $| A_{ii}^{min} | \propto T^{3}$ relation. }
\end{figure} 

 For $t \gg \lambda_{min}^{-1}$, $\delta n(t)$ behaves as $e^{-\lambda_{min} t}$. We now want to look at the behavior of $\lambda_{min}$ as a function of temperature and localization length. Note that the interaction part in the A-matrix (second term in Eq.(\ref{19}(d))) does not contribute much to the eigenvalue distribution at low temperatures as shown in Fig. \ref{fig:3} for all localization lengths considered. In-fact  for $\xi=1$ and $\xi = \,0.5$, the eigenvalue distributions (at low T) are mostly determined by the diagonal part of the $A-matrix$. Consequently, the lowest eigenvalue of the dynamical matrix A ($\lambda_{min}$) approximately equals the smallest value of $A_{ii}$ (defined $A^{min}_{ii}$). In Fig.\ref{fig:4}, we find that $A^{min}_{ii} \propto T^{3}$ for large $\xi$ values. We now propose an argument for this behavior: 

\vspace{2mm}
\textbf{\underline{{$A^{min}_{ii} \propto T^{3}$:}}}
\vspace*{3mm}

 Using Eq.(\ref{19})c, we calculate the $A_{ii}$'s and find that they are smallest for sites around the Fermi level ($E_{i} \simeq \mu$ and so $f_{i}\approx 0.5$), which allows us to consider Eq.(\ref{19})c in the form:
\ba
\label{eq21}
A^{min}_{ii}  &=& -4 \sum_{k \neq i}\, \Gamma_{i k} \nonumber \\
&=& -4 \sum_{r} \sum_{E_{electron}} e^{-r/\xi} \, e^{-\beta |E|}\, F(r,E) \, .
\ea 
Here $F(r,E)$ is the probability of finding an electron or hole having the Hartree energy $E$ at a distance $r$ from a site $i$. Since $\xi$ is large, the electrons will hop to a site so as to minimize the factor ($r/\xi \ + \ \beta |E|$). This means that hops to $r>>1$ are possible. The above Eq.(\ref{eq21}) can now be estimated by
\ba
\label{eq22}
A^{min}_{ii} \propto -4 \times \, \sum_{r} e^{-r/\xi} \, \int^{E_{max}}_{0} g(E)\, e^{-\beta |E|} \, dE \, ,
\ea
where $g(E)$ is the density of states (DOS) of single-particle Hartree energies $(E)$. As discussed earlier, our results (see Fig.\ref{fig:1}(b)) shows that
$g(E) \propto E^{2.0}$. Substituting that into Eq.(\ref{eq22}) we get
\ba
\label{B}
A^{min}_{ii}  \propto T^{3} \, .
\ea
\begin{figure}
\vspace*{1cm}
	\includegraphics[scale=0.3]{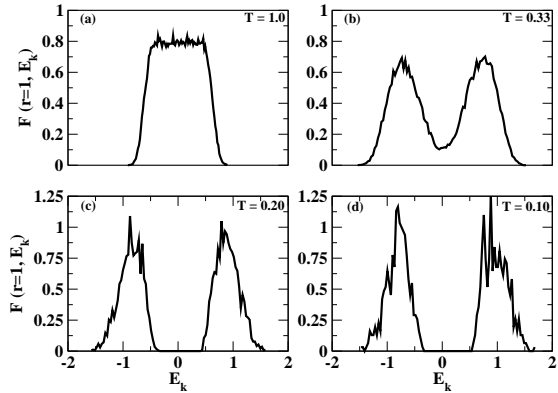}
	\caption{\label{fig:6} (Colour online) (a)-(d) Distribution of the Hartree energy on site $k$, $E_{k}$ (where k are the 6 nearest neighbor sites of $i$) at different temperatures, when the Hartree energy on site $i$, $E_{i}$ are chosen from the interval $[-0.1,0.0]$. }
\end{figure}
 
\begin{figure}
\centering
\subfigure{\includegraphics[width=4cm,angle=-90]{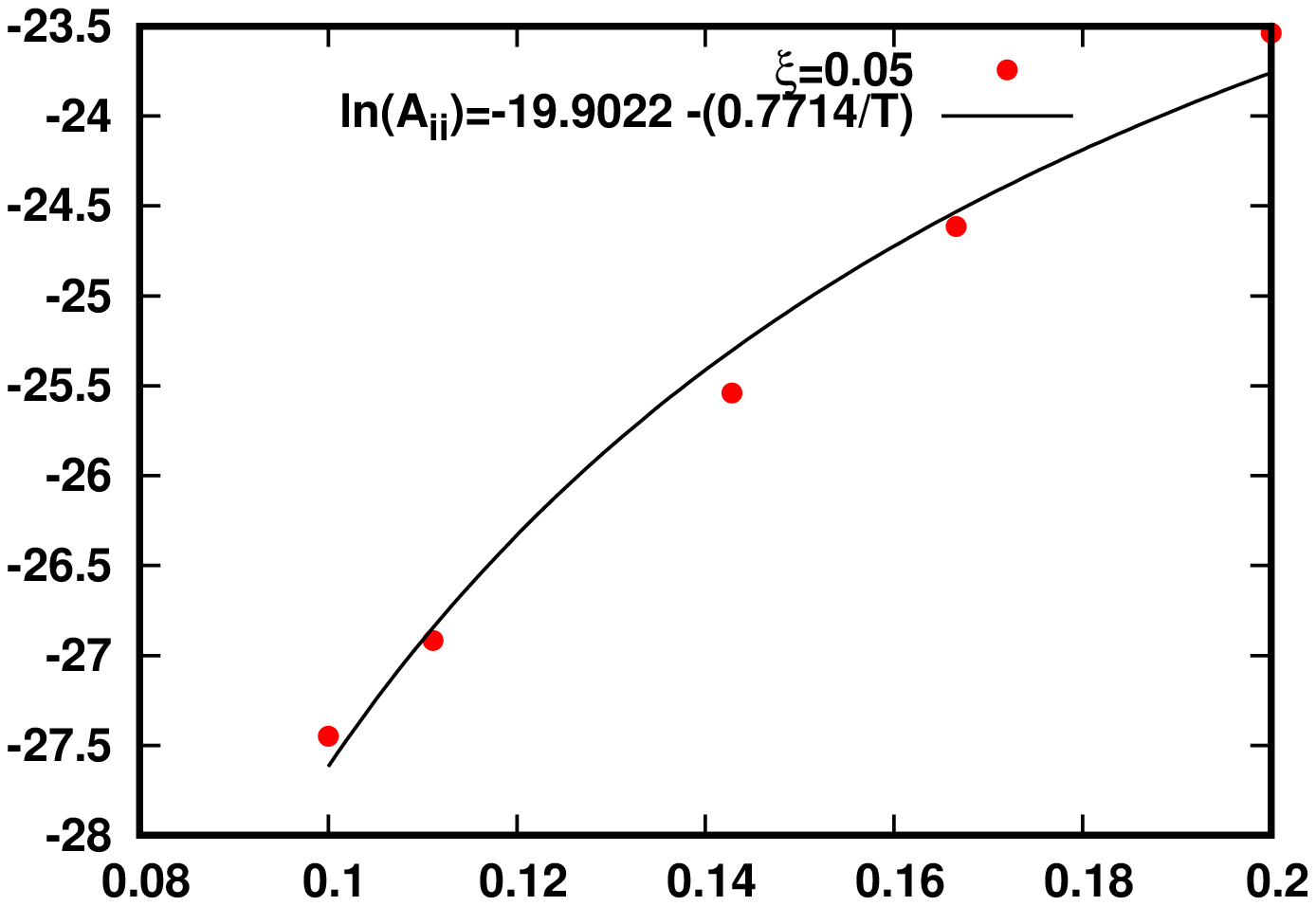}}
\subfigure{\includegraphics[width=4cm,angle=-90]{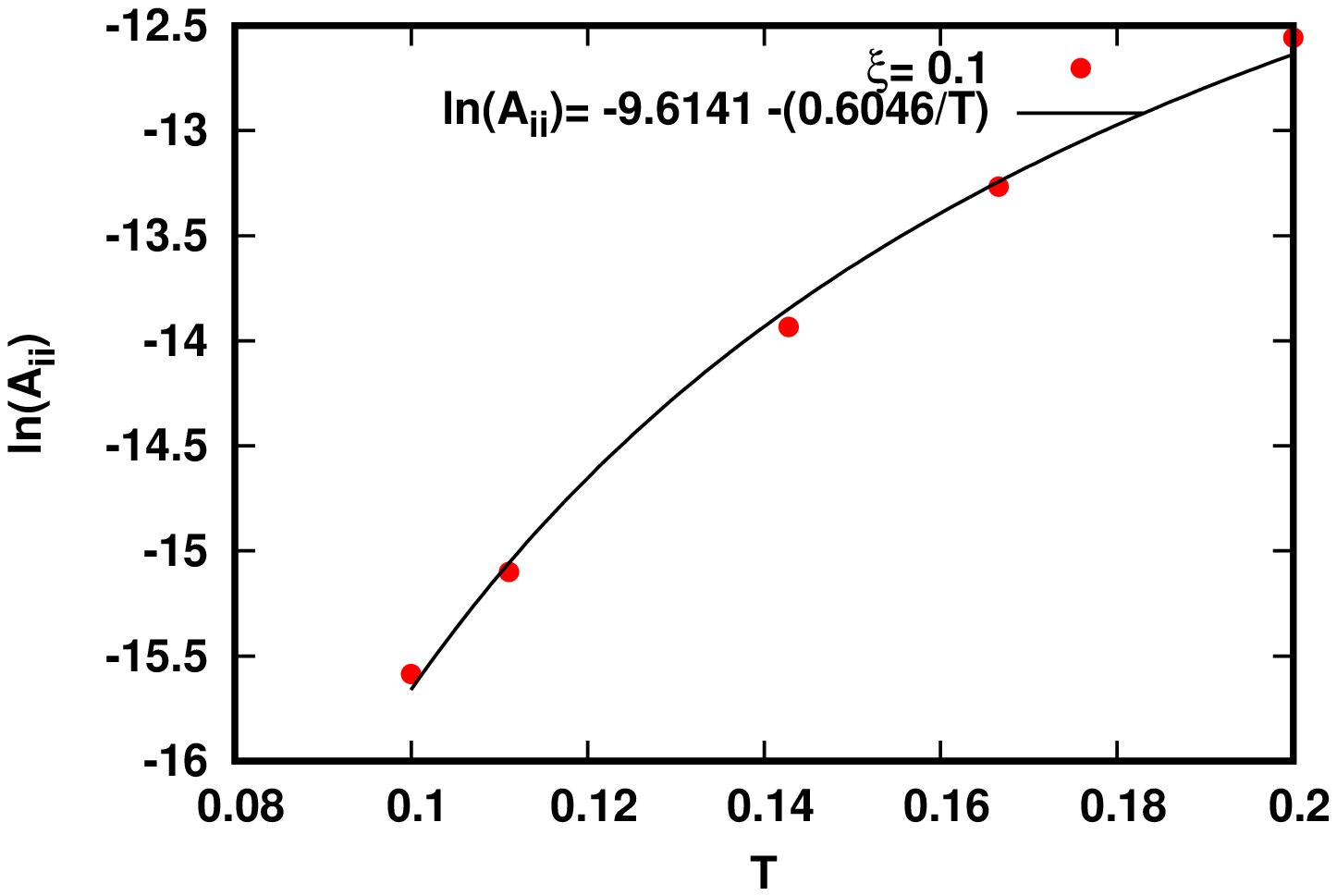}}
\subfigure{\includegraphics[width=4cm,angle=-90]{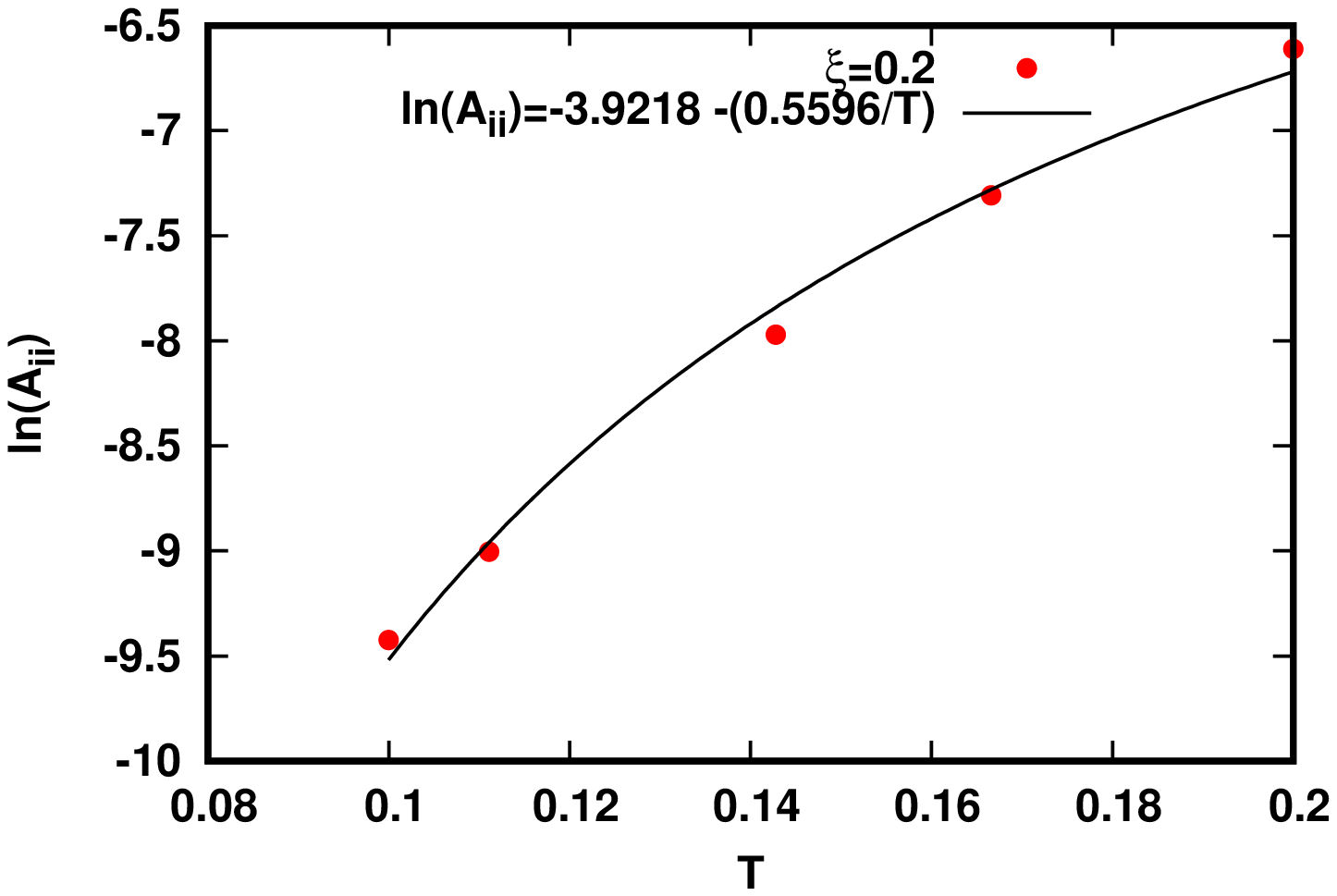}}
\subfigure{}
\caption{\label{fig:5} (Colour online) Lowest value of the diagonal part of $A-matrix$ at $\xi = 0.05,0.1,0.2$ calculated using Eq.(\ref{19}(c)), which is averaged over 100 configurations are plotted against temperature. The solid line shows that the data fits well with $ln(| A_{ii}^{min}) | = a + \frac{b}{T}$ relation for small $\xi$ values. }
\end{figure} 

\begin{figure}
\centering
\subfigure{\includegraphics[width=5cm,angle=-90]{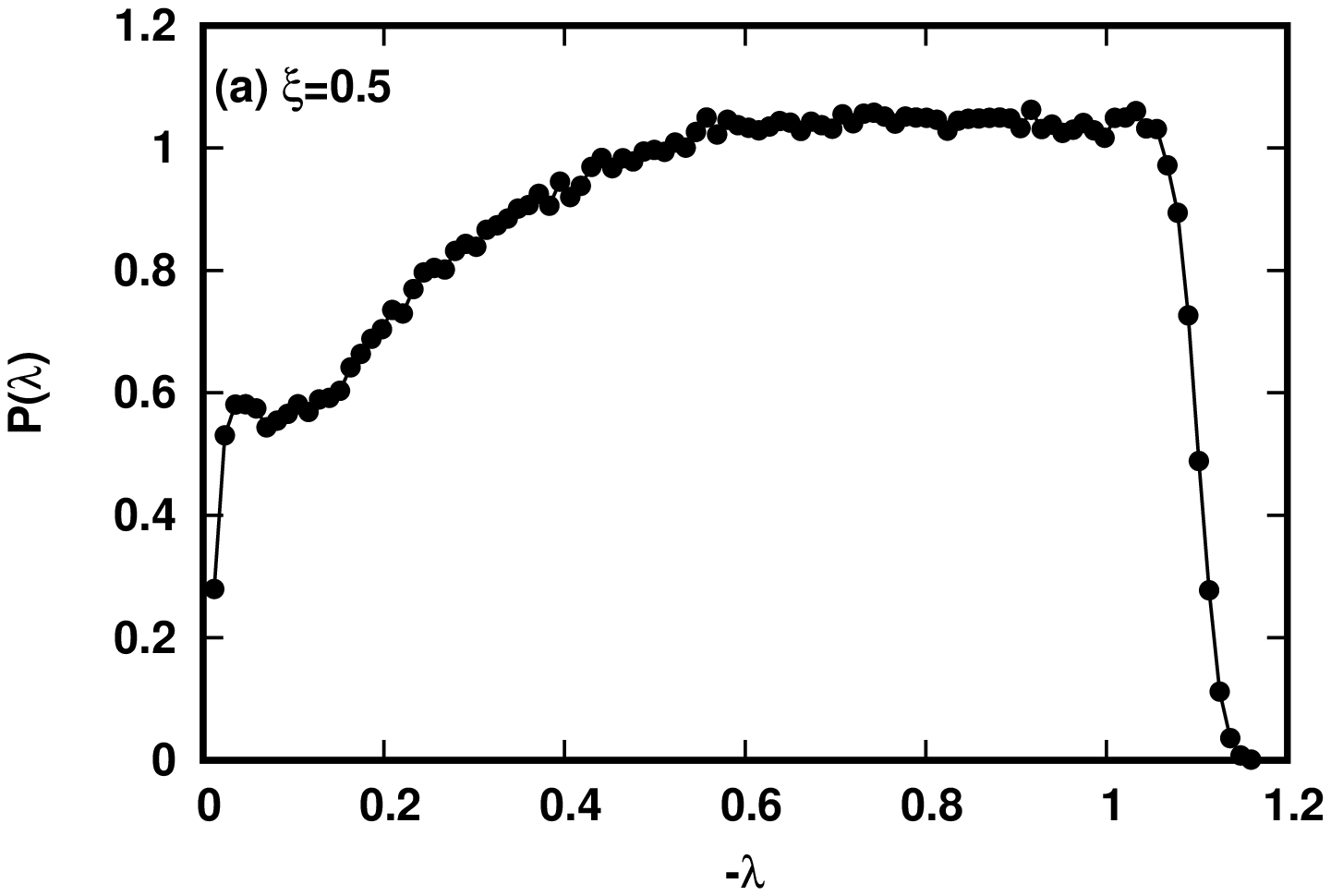}}
\subfigure{\includegraphics[width=5cm,angle=-90]{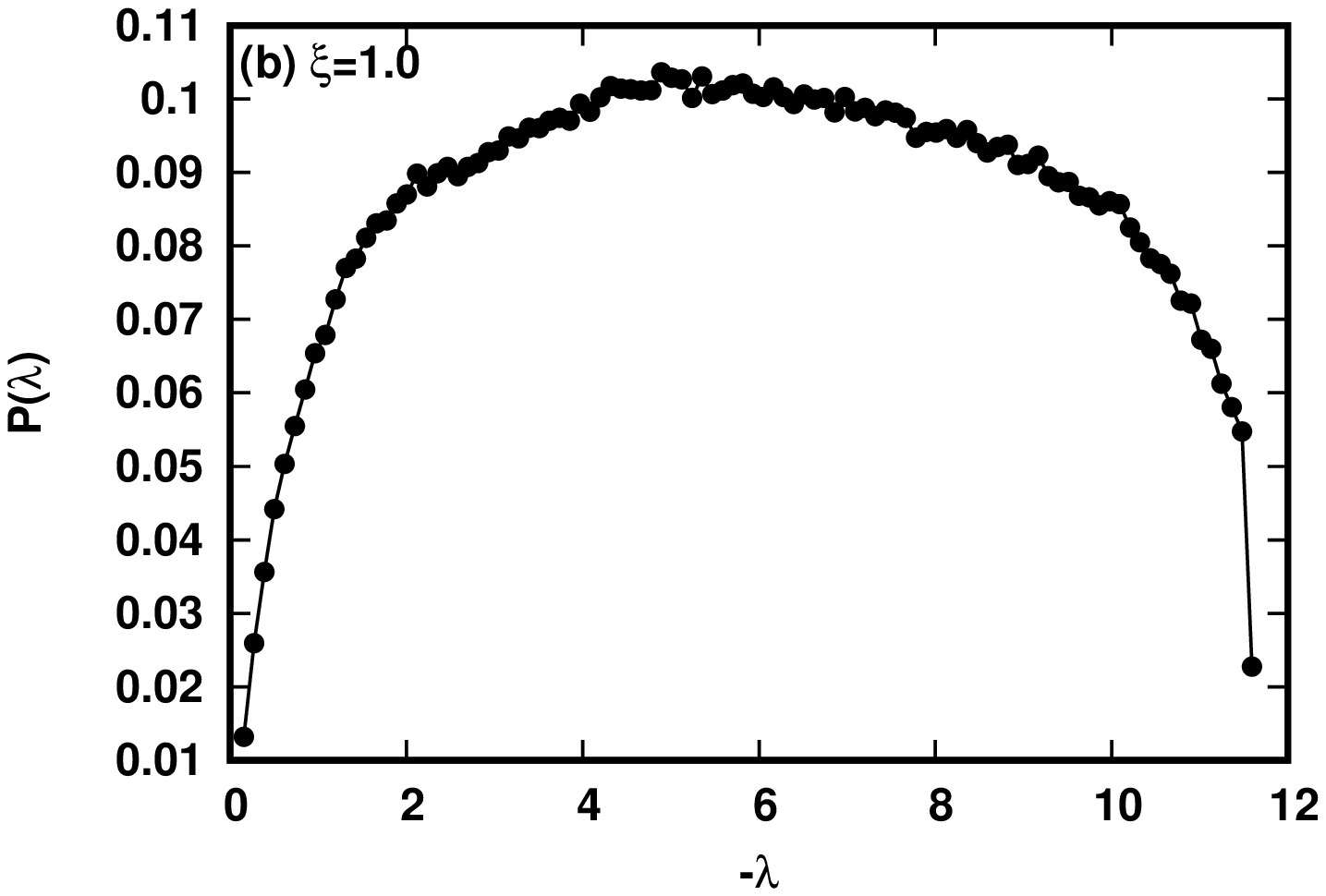}}

\caption{\label{fig:9} Distribution of the eigenvalues of the dynamical matrix A obtained by solving Eq.(\ref{19}), for large localization lengths at $T = 0.1$.}
\end{figure}
  
We now look at the behavior of low temperature $\lambda_{min}$ values at small localization lengths ($\xi=0.2,0.1$,$0.05$). In this case, firstly the $A_{ii}$ distribution is different from the eigenvalue distribution, but the minimum value remains almost the same. More importantly, one should note that the above arguments for $A^{min}_{ii}$ for the temperature range considered $(T=0.1-0.2)$ does not work well when the localization length is very small. Specifically, for small localization lengths the major contribution to $A^{min}_{ii}$ comes from the nearest neighbor sites only (i.e. $r= 1$). So one has to find $F(r=1,E)$ which is the two particle nearest neighbor DOS and insert it into Eq.(\ref{eq21}). In Fig.\ref{fig:6} we show $F(r=1,E)$ at different temperatures for $0 < E_{i} \leq -0.1$. Unlike the full DOS plotted in Fig.\ref{fig:1}(a), there is a hard gap in $F(r=1,E)$ for small energy electrons at low temperatures. This is not surprising, since if one was working with true ground state, then there is a hard gap extending to E $\approx$ 1 in $F(r=1,E)$. The reason behind it is that the ground state is stable against any single electron-hole transition which implies $E_{h} - E_{e} - 1/r_{eh} > 0$. This means that $|E_{h}| + |E_{e}| > 1$ for any nearest neighbor electron-hole pair. This implies that for $\xi \ll 1$ Eq.(\ref{eq21}) reduces to 
\begin{equation}
\label{Aii_min2}
A^{min}_{ii} \approx -e^{-1/\xi} \sum_{j}  e^{-\Delta E_{ij}/T}
\end{equation}
where $\Delta E_{ij}$ is the energy difference between site $i$ and its nearest neighbors $j$. In Fig.\ref{fig:5}, we have shown that $A^{min}_{ii}(T)$ indeed follows the above relation at small localization lengths. Thus, our analysis of the matrix $A_{ii}$ shows that at low temperatures $\lambda_{min}$ obeys different scaling laws for small and large localization lengths.

\begin{figure}
\centering
\subfigure{\includegraphics[width=2.8cm,angle=-90]{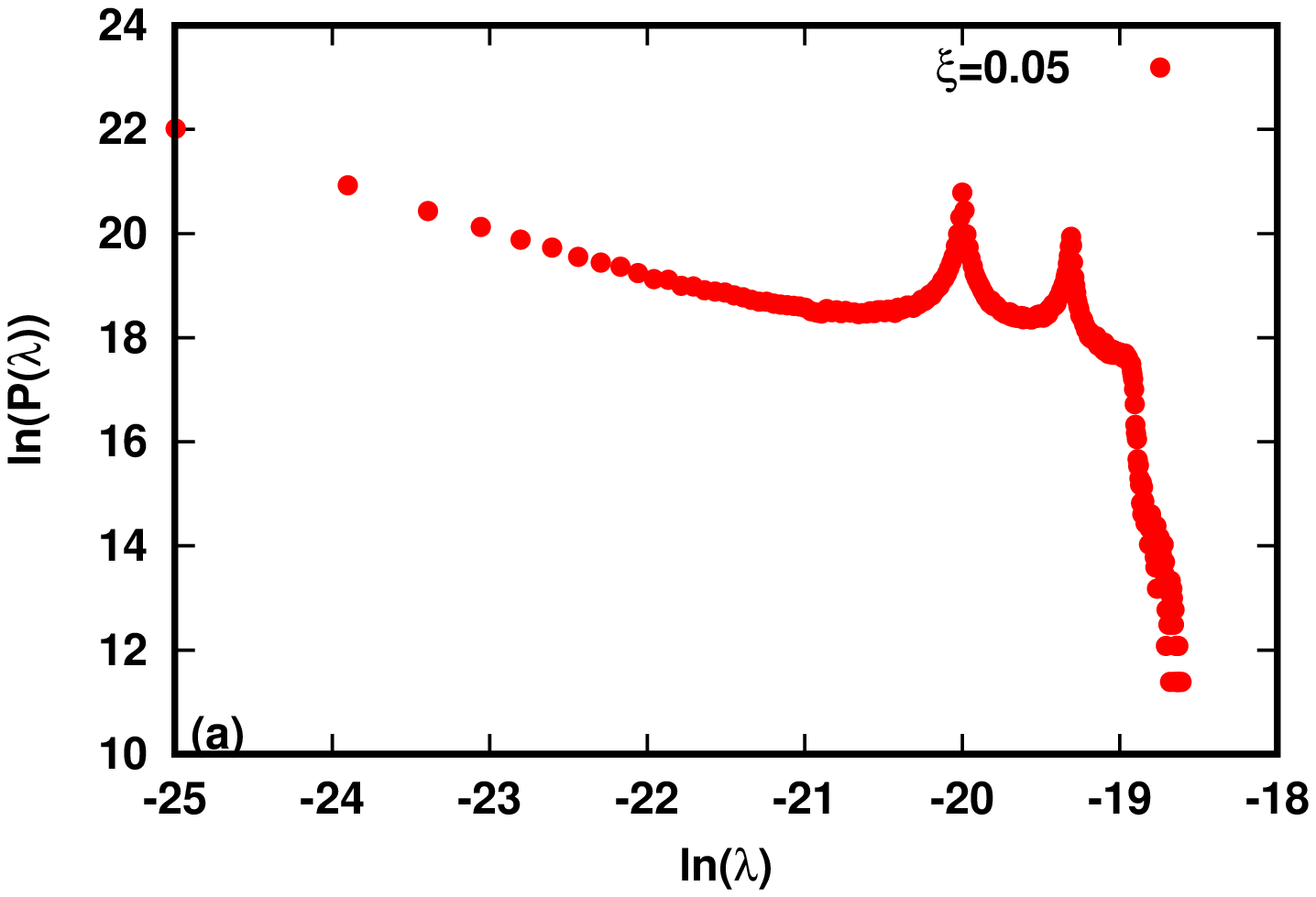}}
\subfigure{\includegraphics[width=2.8cm,angle=-90]{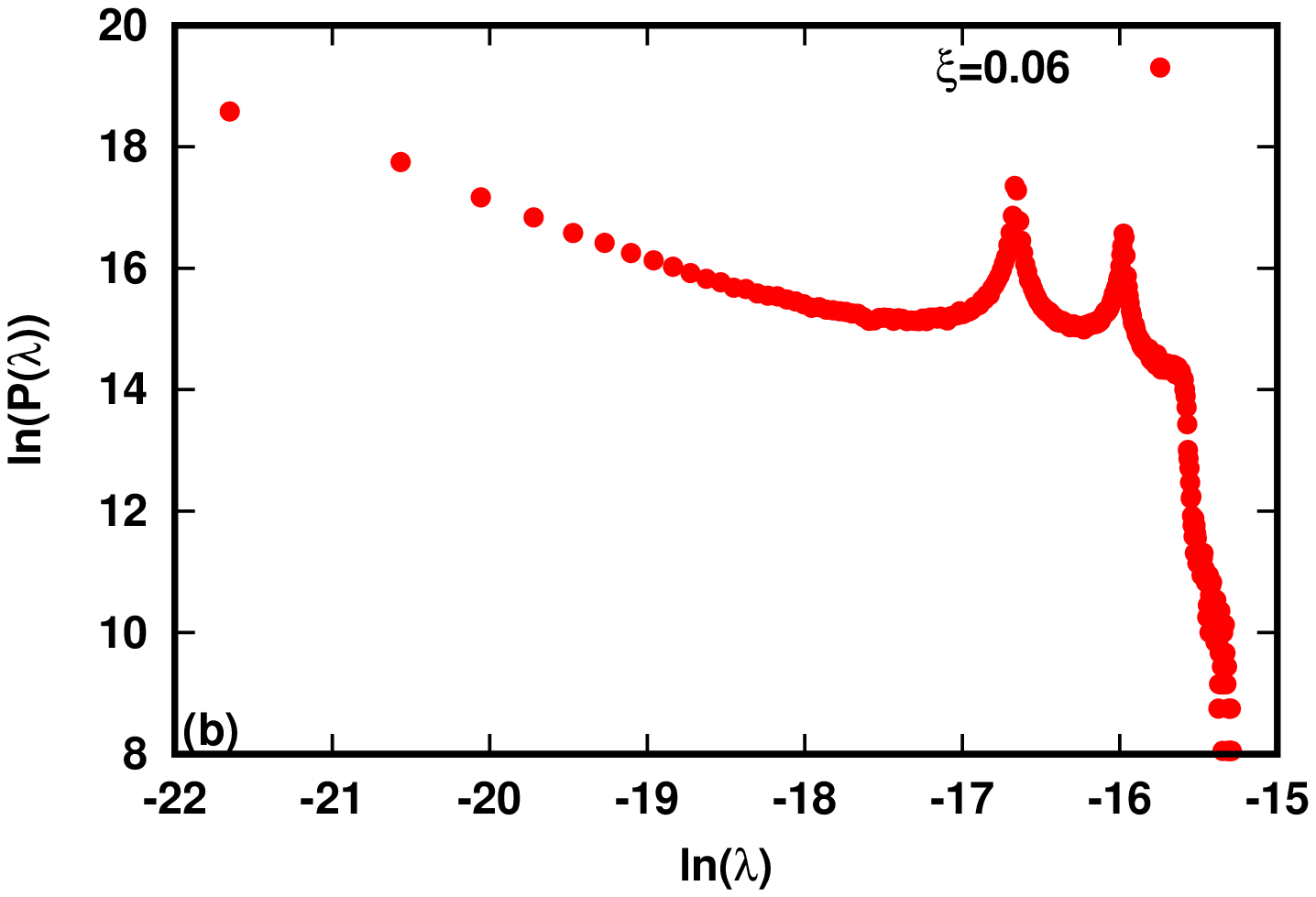}}
\subfigure{\includegraphics[width=2.8cm,angle=-90]{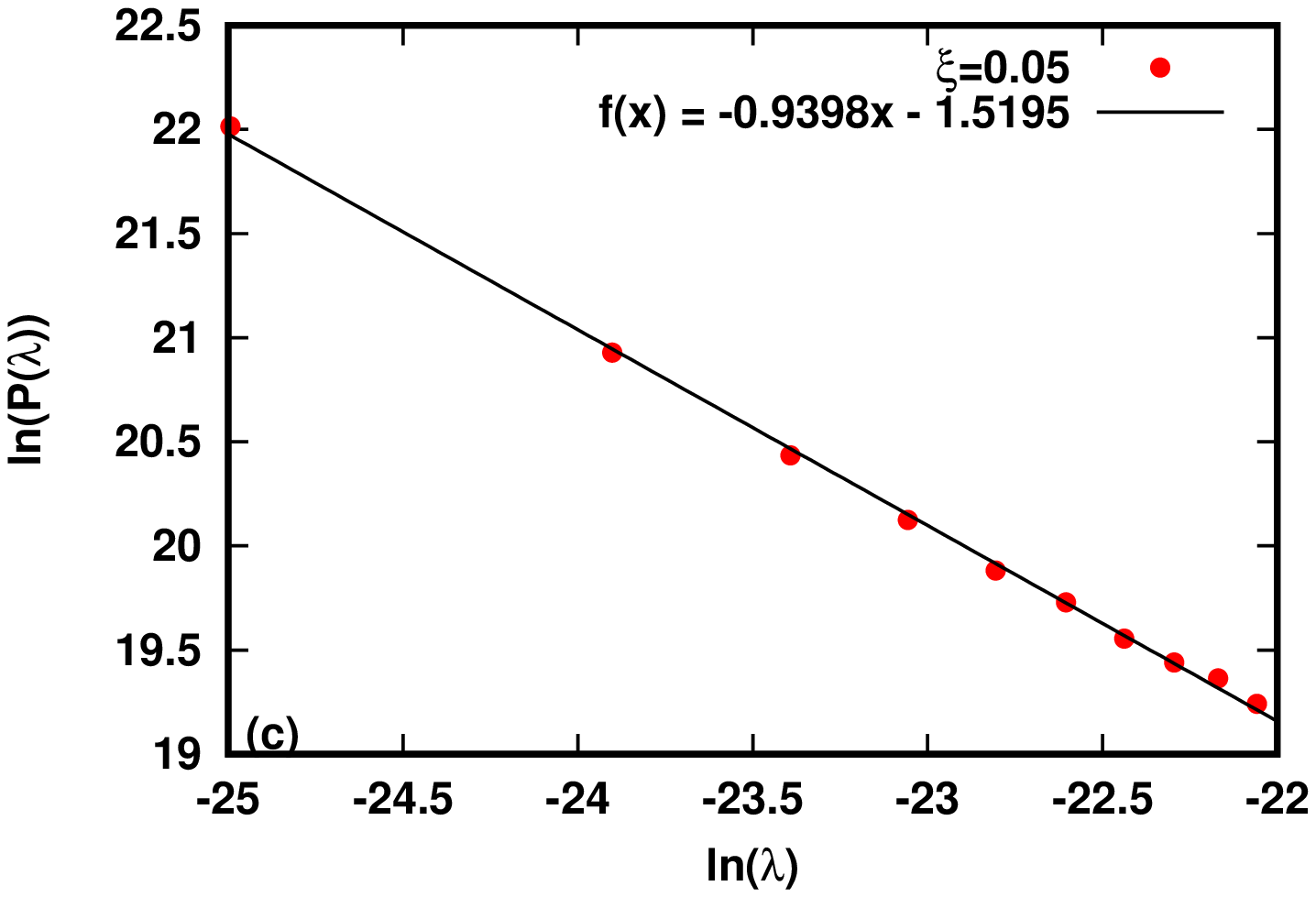}}
\subfigure{\includegraphics[width=2.8cm,angle=-90]{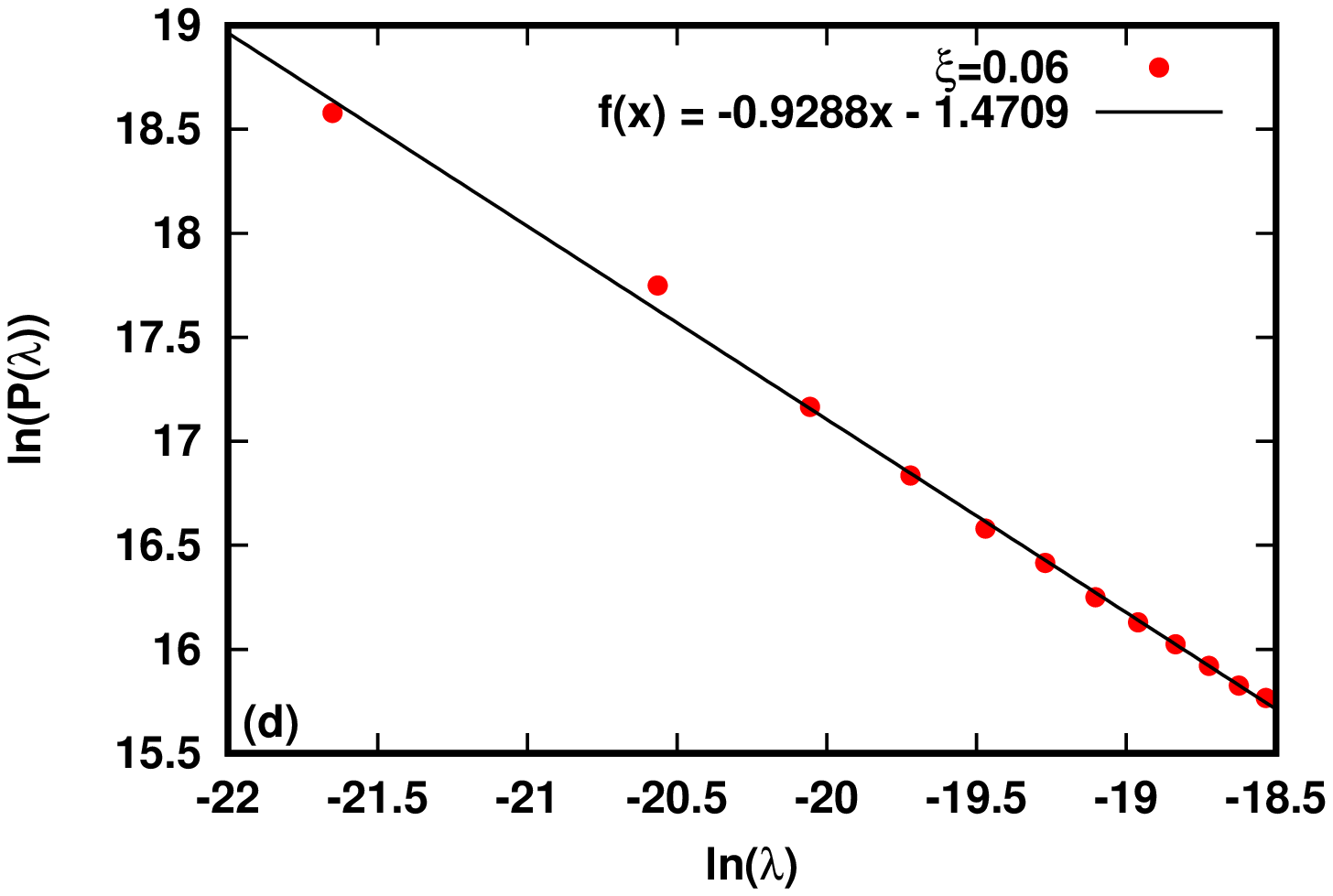}}
\caption{\label{fig:10} (a)-(b) Log-log plot of the distribution of the eigenvalues of the dynamical matrix A obtained by solving Eq.(\ref{19}), for small localization lengths at $T = 0.1$. (c)-(d) For a certain range of $\lambda$ (see text), $ln(P(\lambda))$ vs $ln(\lambda)$ has a linear fit. }
\end{figure}

 We now look at the behavior of the system for $T=0.1$ at time $t<\frac{1}{\lambda_{min}}$ ($\lambda > \lambda_{min}$) for different localization lengths. When the localization length is large ($\xi=0.5,1.0$) we find that $P(\lambda)$ is almost flat. This is shown in Fig.{\ref{fig:9}}. This implies an exponential decay for $\delta n(t)$. $\delta n(t) \sim exp(-\lambda_{1}t)$, where $\lambda_{1}$ is the smallest eigenvalue at which the flat region starts. For small localization lengths ($\xi=0.05,0.06$), the variation of $P(\lambda)$ vs $\lambda$ are shown in Fig.{\ref{fig:10}(a-b)}. For $\xi=0.05$, one sees sharp peaks at $ln(\lambda)=-20$ and $-19.3$. Since $\xi$ is small, the relaxation is dominated by the nearest neighbor hopping. For all sites for which $k \neq 0$ nearest neighbor hops with a decrease in energy are available, $A_{ii}$ is given by 
\begin{equation}
\label{Aeq}
A_{ii}=k \,exp\bigg( -\frac{1}{\xi} \bigg) \, ,
\end{equation}
For $\xi=0.05$ we find $A_{ii}=exp(-20)$ for $k=1$ and $A_{ii}=exp(-19.3)$ for $k=2$, which correspond to the peaks at $ln(\lambda)=-20$ and $-19.3$ respectively in Fig.\ref{fig:10}(a). Similar behavior was seen at $\xi=0.06$ as shown in Fig.\ref{fig:10}(b). So at short times, $\delta n(t)$ will decay according to $\sum_{p} \,  e^{-\lambda_{p}t}$, where $\lambda_{p}$ correspond to eigenvalues at which $P(\lambda)$ has peaks.

When nearest neighbor hops, which lead to decrease in energy ($\Delta E \leq 0$) are not possible, one would get a transition to nearest neighbor site with $\Delta E > 0$. In this case $A_{ii}$ can be written as 
\begin{equation}
\label{Aeq2}
A_{ii} = \sum_{j} e^{-r_{ij}/\xi} \, e^{-\Delta E/T} \, .
\end{equation}
For the $\lambda-$range considered in Fig.\ref{fig:10}(c-d), we find that $A_{ii} \sim \lambda$. So, using Eq.(\ref{Aeq2}) 
\begin{equation}
\label{Peq}
P(\lambda) = \int_{\Delta E_{min}}^{\Delta E_{max}} \delta(\lambda - ce^{-\Delta E/T}) \,P(\Delta E) \,d(\Delta E)
\end{equation}
where $c=e^{-1/\xi}$, $\Delta E_{min} \approx 2T$ and $\Delta E_{max} \approx 5T$. Approximating $P(\Delta E)$ (for $\Delta E_{min} < \Delta E < \Delta E_{max}$) by a uniform distribution on gets
\begin{equation}
\label{logP}
P(\lambda) \sim \frac{1}{\lambda} \, .
\end{equation}
In Fig.\ref{fig:10}(c-d), we plot $P(\lambda) \,vs \, \lambda$ for the regime where nearest neighbor activated hoping takes place. In this regime we find $P(\lambda) \sim 1/\lambda$. This leads to logarithmic temporal dependence of the relaxation for intermediate times. Recently, a crossover from logarithmic time dependence to an exponential dependence was shown in  a non-equilibrium study \cite{z18} of excess conduction $\Delta G(t)$ in disordered indium oxide. In this it was shown that as one approaches metal insulator transition from the insulating side, the crossover time becomes smaller. This implies that as the disorder in the system decreases and localization length increases the crossover time to exponential decay decreases. Since $A_{ii}^{min}$ is equal to $\lambda_{min}$, Eq.(\ref{Aii_min2}) shows that the $\lambda_{min}$ increases as localization length increases. 
This implies that $\tau_{max} = 1/\lambda_{min}$ will decrease as the localization length increases and crossover from the logarithmic behavior to exponential decay happens faster.   

\begin{figure*}
\centering
\subfigure{\includegraphics[width=3.5cm,angle=-90]{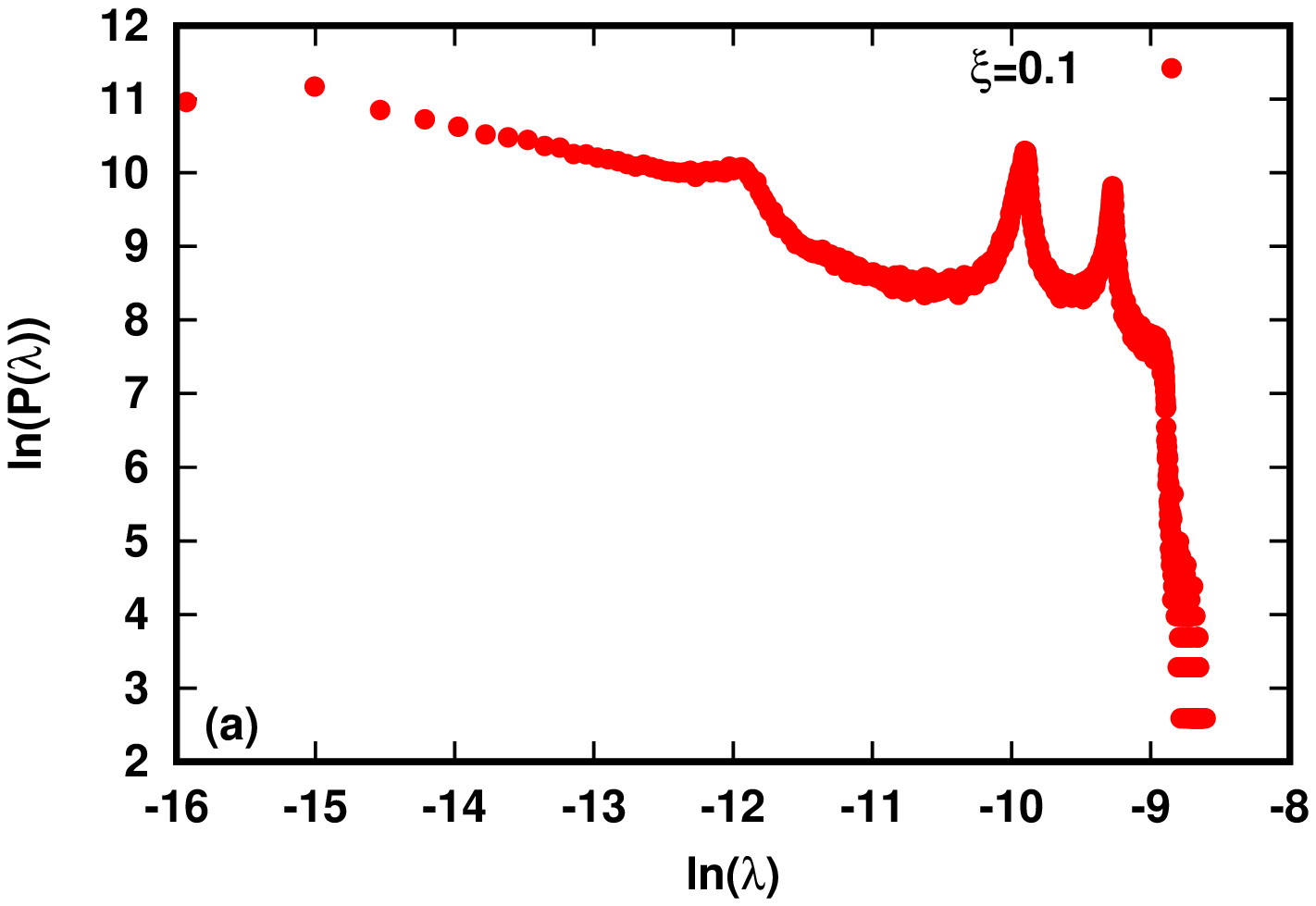}}
\subfigure{\includegraphics[width=3.5cm,angle=-90]{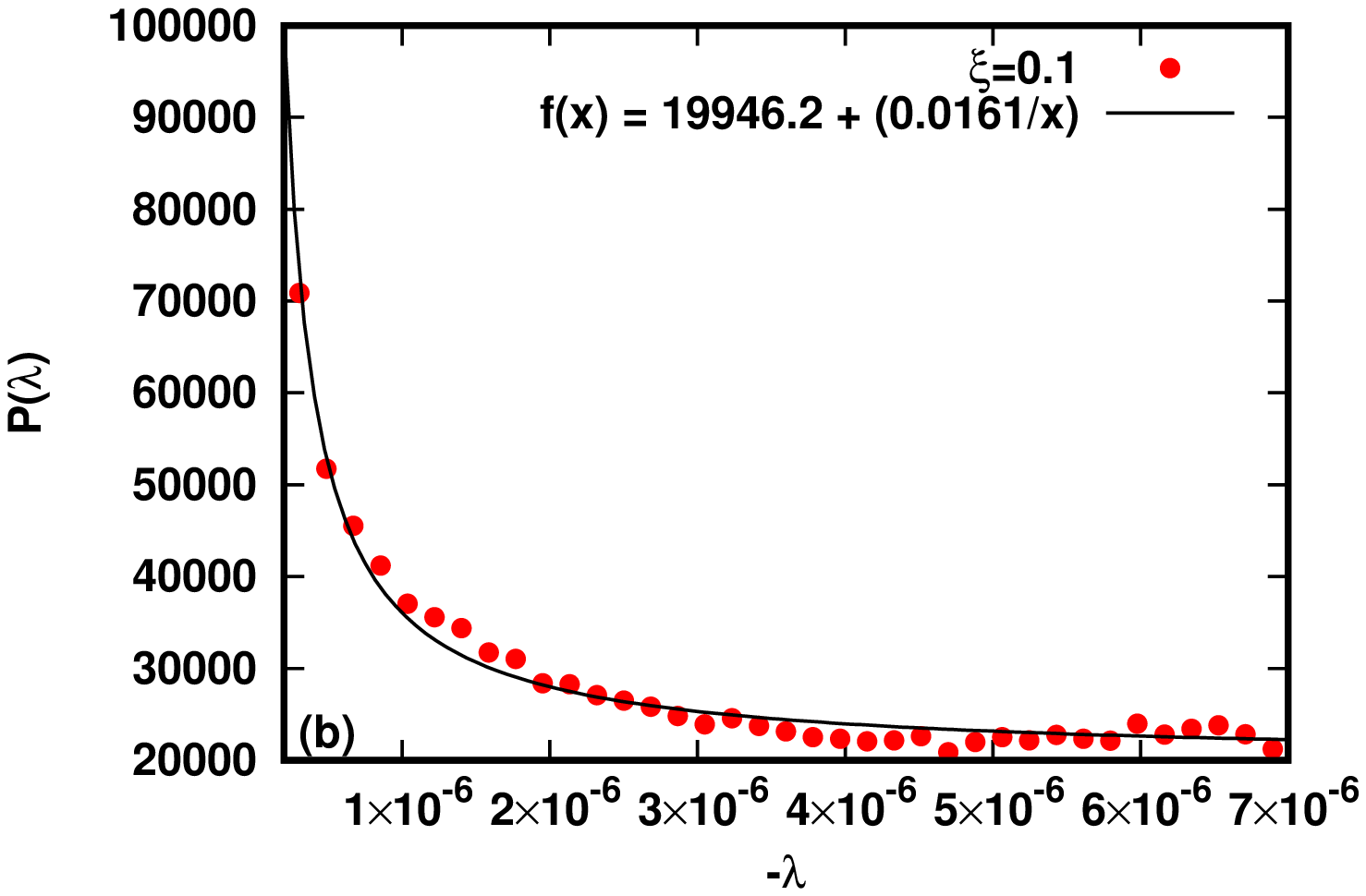}}
\subfigure{\includegraphics[width=3.5cm,angle=-90]{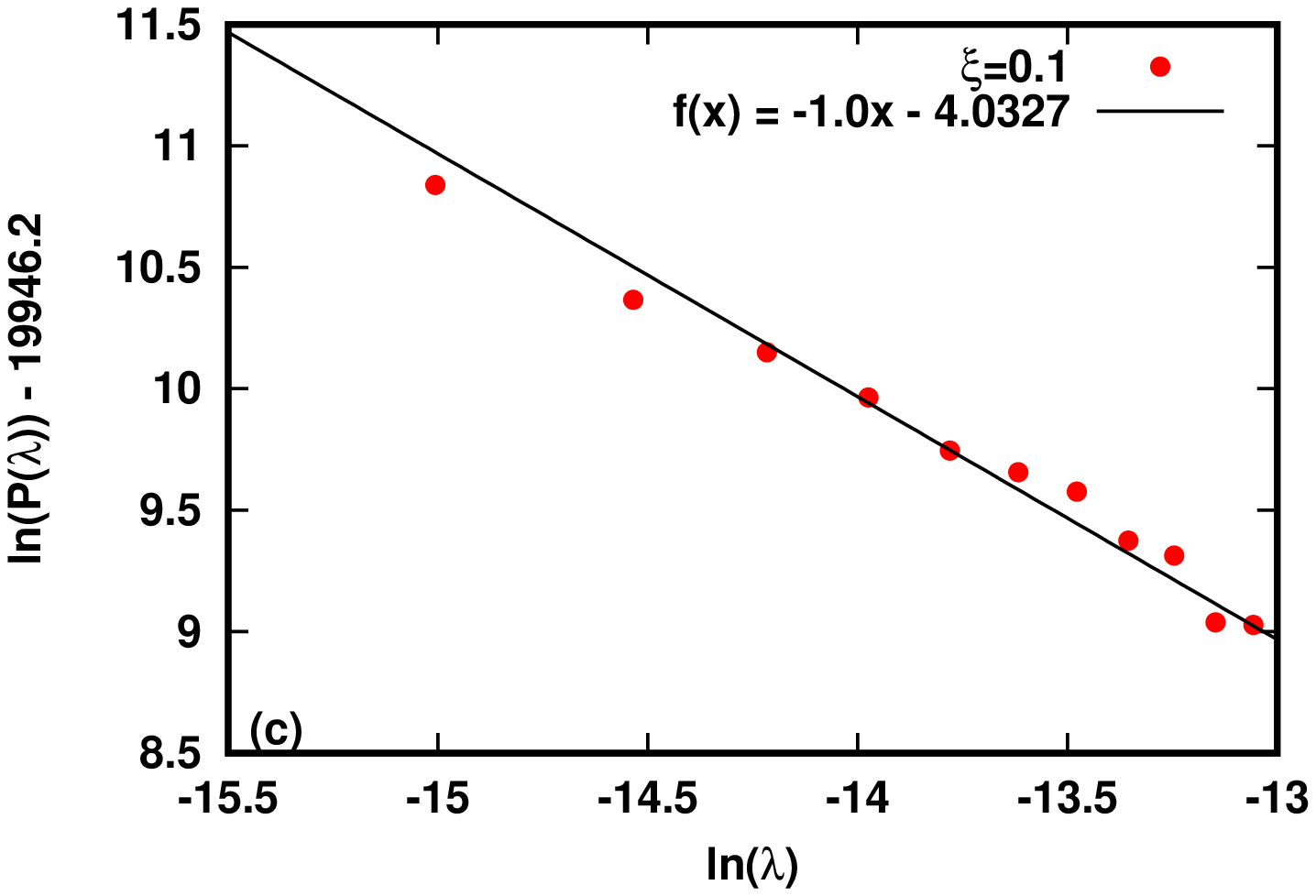}}
\subfigure{\includegraphics[width=3.5cm,angle=-90]{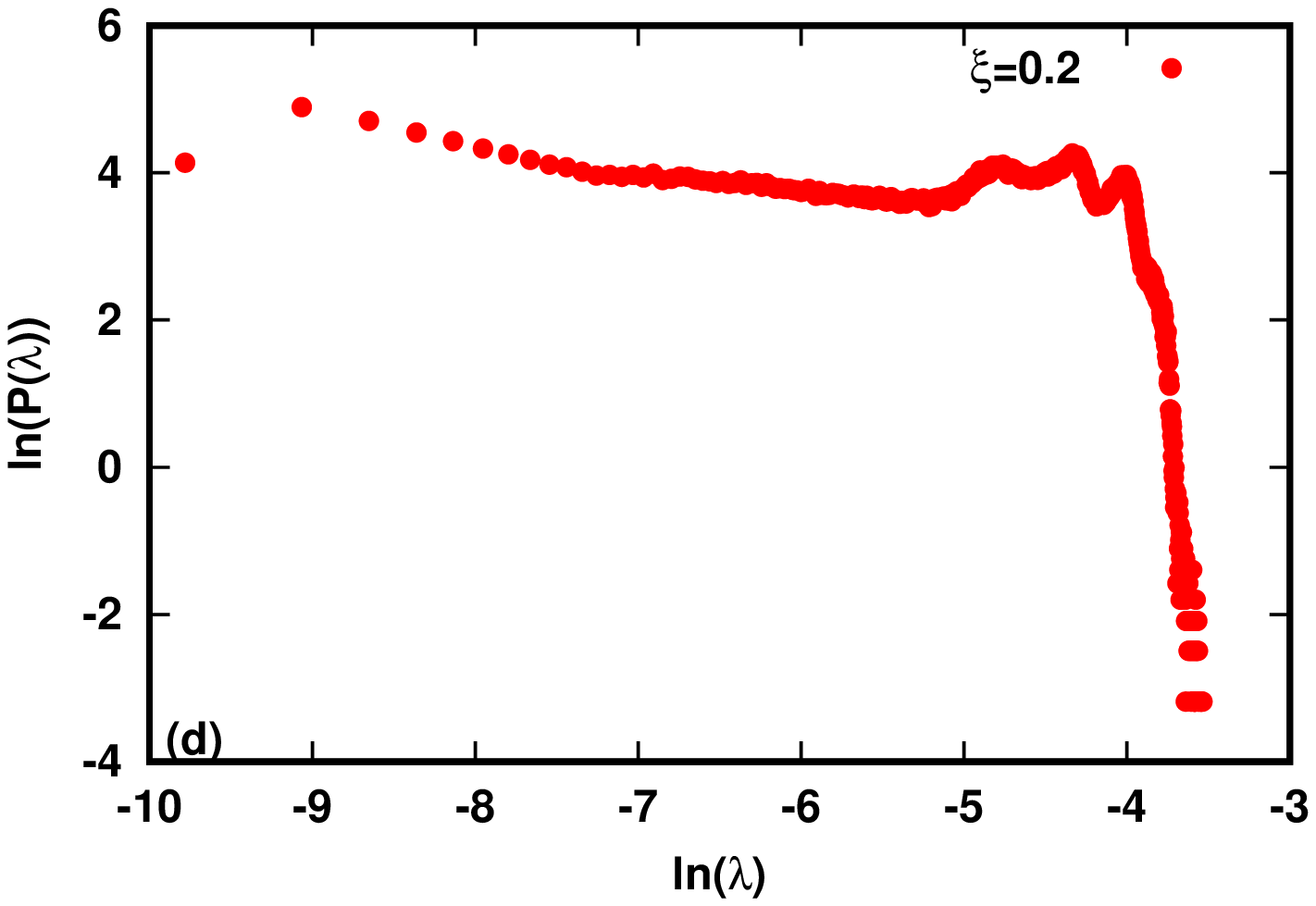}}
\subfigure{\includegraphics[width=3.5cm,angle=-90]{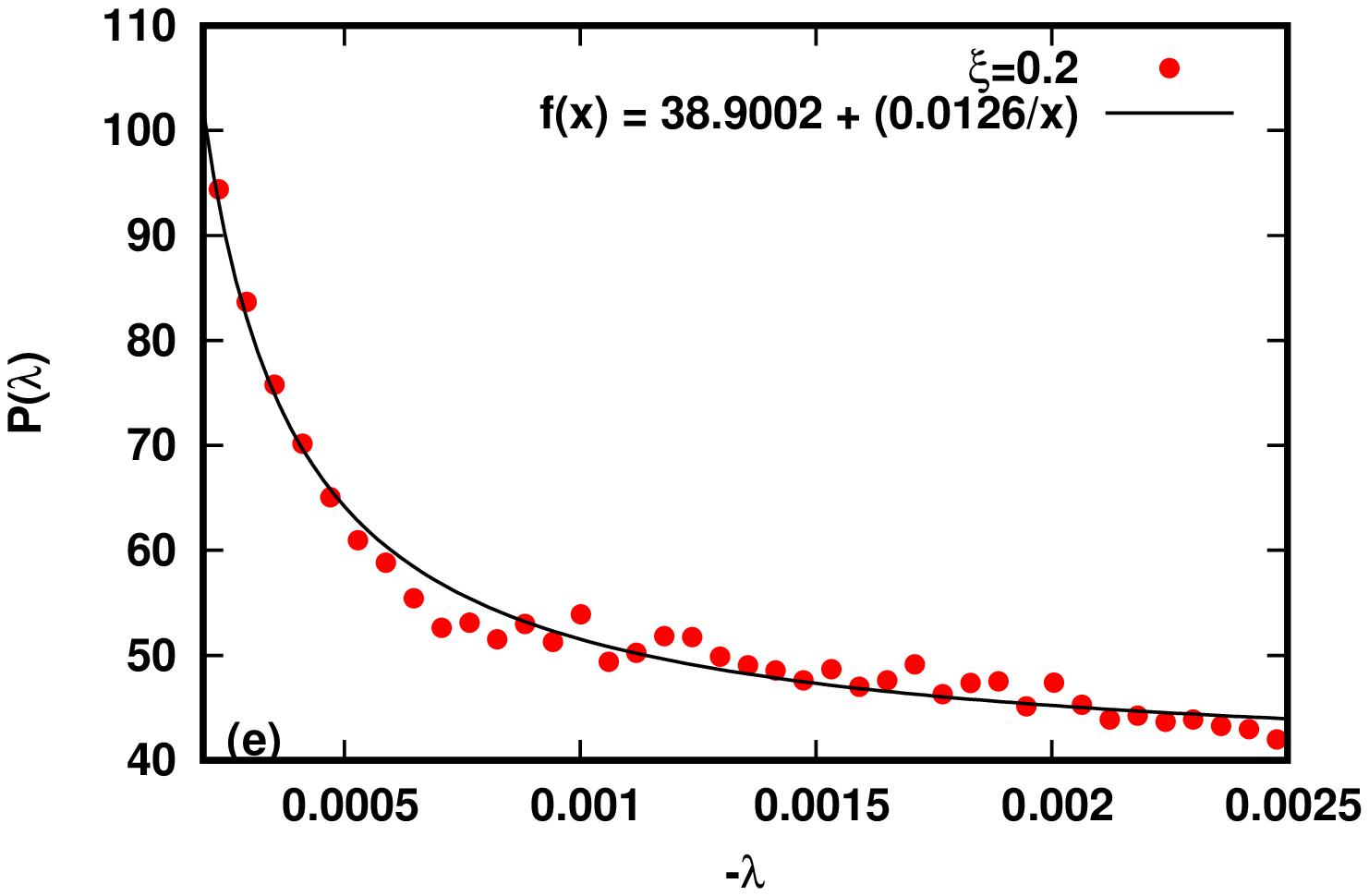}}
\subfigure{\includegraphics[width=3.5cm,angle=-90]{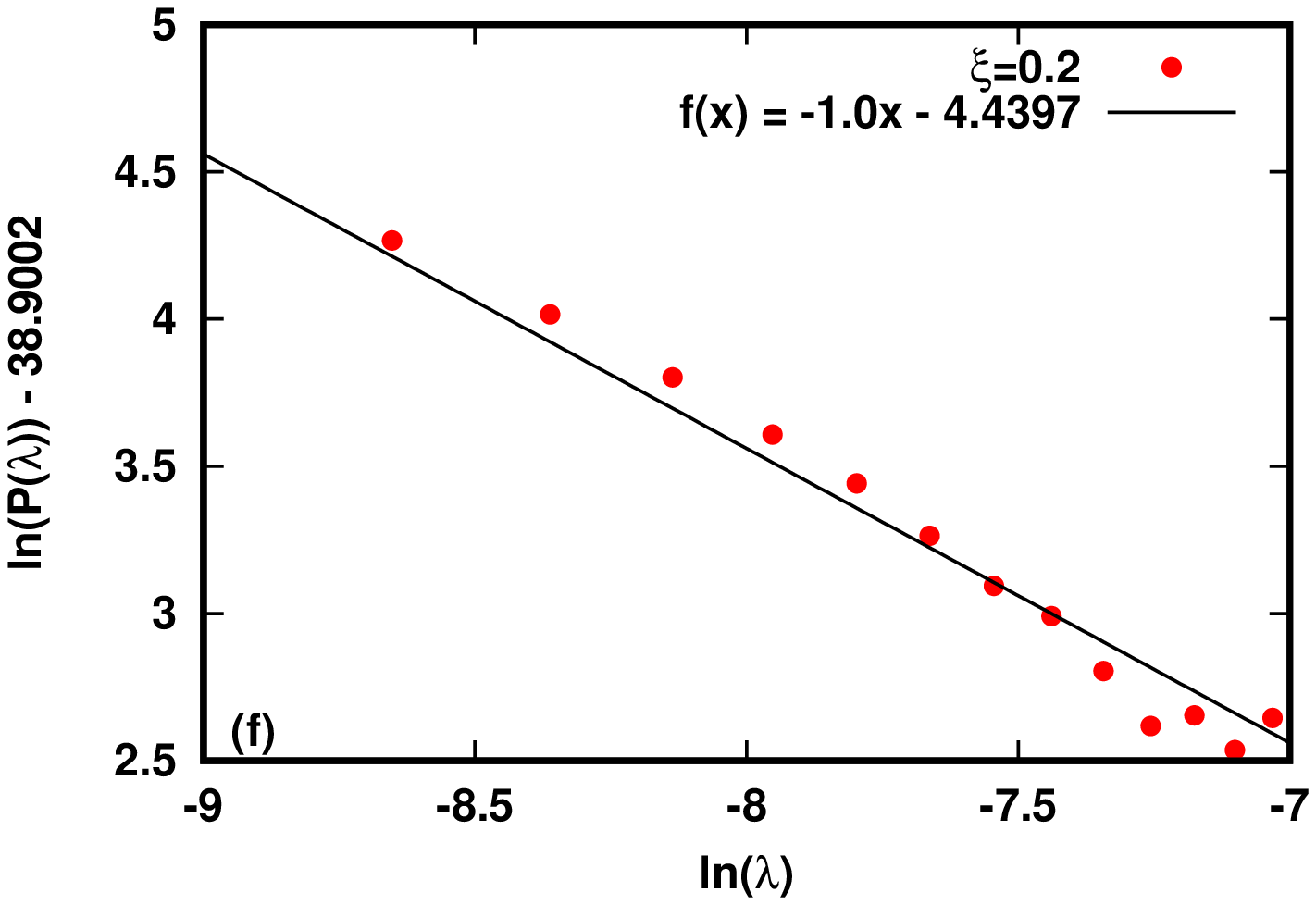}}
\caption{\label{fig:11} (a),(d) Log-log plot of the distribution of the eigenvalues of the dynamical matrix A obtained by solving Eq.(\ref{19}), for intermediate localization lengths at $T = 0.1$. (b),(e) Behavior of $P(\lambda)$ at intermediate times. The solid lines shows that the data fits well with $P(\lambda) \propto a + \frac{b}{\lambda}$ relation. (c),(f) The fit to the logarithmic distribution after subtracting the background constant is close to linear.}
\end{figure*}

At intermediate localization lengths ($\xi=0.1,0.2$) one sees that for large $\lambda$'s, there are peaks corresponding to next nearest neighbor hops with decrease in energy. For $\lambda =$ -15 to -13 at $\xi=0.1$ and $\lambda =$ -9 to -7 at $\xi=0.2$, $ln(P(\lambda))$ vs $ln(\lambda)$ has a linear fit but the slope is not equal to $-1$. The reason is that for intermediate $\xi$'s there is contribution to $\lambda$ from next nearest neighbor hops as well as nearest neighbor hops. In Fig.\ref{fig:11}(b,e) we have plotted $P(\lambda)$ vs $\lambda$ for these regions. We get $P(\lambda) = a + b/\lambda$ with value of $a \gg b$ for both $\xi = 0.1$ and $\xi = 0.2$. This form of $P(\lambda)$ implies relaxation behavior of the form $a \, e^{-\lambda_{1}t} + b \,ln(t)$ where ($\lambda_{1}$ is the minimum value of $\lambda$ for the range under consideration). Since $a \gg b$, it is quite possible that exponential decay will overshadow the logarithmic decay in relaxation of $\delta n(t)$.

\section{\label{sec:level4}Summary}
We consider here the relaxation properties of the three-dimensional Coulomb Glass lattice model in which all the electron states are localized and the dynamics occurs through phonon-assisted hopping among these states. The master equation governing the dynamics of the system is approximated via mean-field theory.

The relaxation law for a range of localization lengths is studied here. The dependence of the relaxation on the
localization length can be summarized as follows: 

(i) For small localization length $\xi \ll 1$, near neighbor hopping is strongly dominant. This results in $P(\lambda) \sim 1/\lambda$ distribution for small $\lambda$'s which leads to a logarithmic temporal dependence of the relaxation at intermediate times.

(ii) For intermediate $\xi$ values (0.1,0.2) next nearest neighbor (n.n.n) contribution also becomes important in relaxation. We find that the relaxation is not purely logarithmic at intermediate times ($P(\lambda) \propto a + b/\lambda$, $a \gg b$).

(iii) For larger values ($\xi=0.5$ and larger) the system relaxes by performing hops to all distances, and no $1/\lambda$ distribution is seen, consequently no logarithmic temporal dependence.

 Finally, we looked at relaxation for times $t \gg \lambda_{min}^{-1}$. We have found that although the full eigenvalue distribution is not much affected by the Coulomb interaction term in the linear dynamical matrix, one can gain a better understanding of the behavior of low-temperature dynamics by looking at the role of the gap in the density of states in the decay process and the range of hopping. The gap in the density of states exists due to the long-range nature of Coulomb interactions and so the interactions play an important role in the relaxation process. For small localization lengths one finds that the $\lambda_{min} \propto e^{-c/T}$ where $c$ is a constant. This implies that time at which exponential decay starts increases exponentially with a decrease in temperature. This may explain why the transition from logarithmic decay to exponential decay is not seen in most experiments.
 
 Recently, a non-equilibrium study \cite{z18} of excess conduction $\Delta G(t)$ in disordered indium oxide showed a crossover from logarithmic time dependence to an exponential dependence. The crossover time became smaller as the metal insulator transition was approached from the insulating side. This implies that as the disorder in the system decreases and localization length increases the crossover time to exponential decay decreases. In our formalism, the crossover time is $1/\lambda_{min}$ which also decreases with increase in localization length. 
 
 Further work is required to establish results in the case where the distance between sites is a continuous variable and not a discrete value (as was the case in the present work).

\section*{ACKNOWLEDGEMENT}
 PB gratefully acknowledges IISER Mohali and the Israel Science Foundation (Grant No. 2300/19) for the financial support. Illuminating discussions with A. Amir and Z. Ovadyahu are gratefully acknowledged. We wish to thank NMEICT cloud service provided by BAADAL team, cloud computing platform, IIT Delhi for the computational facility and Ben Gurion
University of the Negev for access to their HPC resources.


\begin{thebibliography}{99}
 	
 \bibitem{jpt82}
  J. H. Davies, P. A. Lee and T. M. Rice, Phys. Rev. Lett. \textbf{49}, 758 (1982).
  
 \bibitem{mbld82}
  M. Gr\"{u}newald, B. Pohlman, L. Schweitzer and D. W\"{u}rtz, J. Phys. C \textbf{15}, L1153 (1982).  

 \bibitem{mm82}
 M. Pollak and M. Ortu\~{n}o, Sol. Energy Mater. \textbf{8}, 81 (1982).
 
 \bibitem{m84}
  M. Pollak, Philos. Mag. B \textbf{50}, 265 (1984).
  
 \bibitem{ec94}
  E. R. Grannan and C. C. Yu, Phys. Rev. Lett. \textbf{73}, 2934 (1994).  	

 \bibitem{ab75}
  A. L. Efros and B. I. Shklovskii, J. Phys. C: Solid State Phys. \textbf{8}, L49 (1975). 
  
 \bibitem{amb92}
  A. M\"{o}bius, M. Richter and B. Dritter, Phys. Rev. B. \textbf{45}, 11568 (1992).  

 \bibitem{pvs17}
  P. Bhandari, V. Malik and S. R. Ahmad, Phys. Rev. B. \textbf{95}, 184203 (2017).
  
 \bibitem{pv17}
  P. Bhandari and V. Malik, J. Phys.: Condens. Matter \textbf{29}, 485402 (2017).  
  
 \bibitem{sabb79}
  S. D. Baranovskii, A. L. Efros, B. L. Gelmont and B. I. Shklovskii, J. Phys. C: Solid State Phys. \textbf{12}, 1023 (1979).
  
 \bibitem{avjmy08}
  A. Glatz, V. M. Vinokur, J. Bergli, M. Kirkengen and Y. M. Galperin, J. Stat. Mech. \textbf{P06006}, (2008).  

 \bibitem{ba84}
  B. I. Shklovskii and A. L. Efros, Electronic Properties of Doped Semiconductors, Heidelberg: Springer, Heidelberg, (1984)

 \bibitem{mma13}
  M. Pollak, M. Ortu\~{n}o, and A. Frydman, The Electron Glass, Cambridge University Press, New York, (2013).

 \bibitem{m70}
  M. Pollak, Discuss. Faraday Soc. \textbf{50}, 13 (1970)

 \bibitem{g71}
  G. Srinivasan, Phys. Rev. B \textbf{4}, 2581 (1971).

 \bibitem{jpt84}
  J. H. Davies, P. A. Lee, and T. M. Rice Phys. Rev. B \textbf{29}, 4260 (1984). 

 \bibitem{jm95}
  J. G. Massey and M. Lee, Phys. Rev. Lett. \textbf{75}, 4266 (1995).

 \bibitem{vjp00}
  V. Y. Butko, J. F. DiTusa and P. W. Adams, Phys. Rev. Lett. \textbf{84}, 1543 (2000).
  
 \bibitem{mi04}
  M. M\"{u}ller and L. B. Ioffe, Phys. Rev. Lett. \textbf{93}, 256403 (2004). 
  
 \bibitem{m68}
  N. F. Mott, J. J. Non-Cryst. Solids \textbf{1}, 1 (1968).

 \bibitem{m69}
  N. F. Mott, Phil. Mag. B \textbf{19}, 835 (1969). 
  
 \bibitem{mm09}
  M. Goethe and M. Palassini, Phys. Rev. Lett. \textbf{103}, 045702 (2009).     
  
 \bibitem{bhtag09}
  B. Surer, H. G. Katzgraber, T. G. Zimanyi, A. B. Allgood and G. Blatter, Phys. Rev. Lett. \textbf{105}, 067205 (2009). 
  
 \bibitem{am10}
  A. M\"{o}bius and M. Richter, Phys. Rev. Lett. \textbf{105}, 039701 (2010).   
  
 \bibitem{ajmh19}
  A. Barzegar, J. C. Anderson, M. Schechter and H. G. Katzgraber, Phys. Rev. B. \textbf{100}, 104418 (2019).
  
 \bibitem{av99}
  A. A. Pastor and V. Dobrosavljevi\'{c}, Phys. Rev. Lett. \textbf{83}, 4642 (1999).  

 \bibitem{sv05}
  S. Pankov and V. Dobrosavljevi\'{c}, Phys. Rev. Lett. \textbf{94}, 046402 (2005).       

 \bibitem{ms07}
  M. M\"{u}ller and S. Pankov, Phys. Rev. B. \textbf{75}, 144201 (2007).  

 \bibitem{am82}
  A. J. Bray and M. A. Moore, J. Phys. C: Solid State Phys. \textbf{15}, 2417 (1982).
      
 \bibitem{l78}
  L. C. E. Struik, \textit{Physical Aging in Amorphous Polymers and Other Materials} (Elsevier, Amsterdam, 1978).
 
 \bibitem{jlj97}
  J. P. Bouchaud, L. F. Cugliandolo, and J. Kurchan, in \textit{Spin
  Glasses and Random Fields}, edited by A. P. Young (World Scientific, Singapore, 1997)

 \bibitem{l02}
  L. F. Cugliandolo, in \textit{Slow Relaxation and Nonequilibrium Dynamics in Condensed Matter}, (Les Houches Session LXXVII, 1-26 July, 2002).

 \bibitem{e97}
  E. Vincent et al., in \textit{Complex Behavior of Glassy Systems}, edited
  by M. Rubi and C. Perez-Vicente (Springer, Berlin, 1997).
    
 \bibitem{lj93}
  L. F. Cugliandolo and J. Kurchan, Phys. Rev. Lett. \textbf{71}, 173
  (1993). 
  
 \bibitem{lj94}
  L. F. Cugliandolo and J. Kurchan, J. Phys. A 27, 5749 (1994).

 \bibitem{azm000}
  A. Vaknin, Z. Ovadyahu and M. Pollak, Phys. Rev. B. \textbf{61}, 6692 (2000).  
  
 \bibitem{add05}
  A. B. Kolton, D. R. Grempel and D. Dom\'{i}nguez, Phys. Rev. B. \textbf{71}, 024206 (2005).
  
 \bibitem{mj09}
  M. Kirkengen and J. Bergli, Phys. Rev. B. \textbf{79}, 075205 (2009).  
 
 \bibitem{ayy09}
  A. Amir, Y. Oreg and Y. Imry, Phys. Rev. Lett. \textbf{103}, 126403 (2009). 
  
 \bibitem{ayy10}
  A. Amir, Y. Oreg and Y. Imry, Phys. Rev. Lett. \textbf{105}, 070601 (2010). 
   
 \bibitem{jy12}
  J. Bergli and Y. M. Galperin, Rev. Cub. Fis. \textbf{29}, 1E9 (2012).

 \bibitem{azm00}
  A. Vaknin, Z. Ovadyahu and M. Pollak, Phys. Rev. Lett. \textbf{84}, 3402 (2000).

 \bibitem{azm02}
  A. Vaknin, Z. Ovadyahu and M. Pollak, Phys. Rev. B. \textbf{65}, 134208 (2002). 
  
 \bibitem{vz04}
  V. Orlyanchik and Z. Ovadyahu, Phys. Rev. Lett. \textbf{92}, 066801 (2004).
          
 \bibitem{mzm93}
  M. Ben-Chorin, Z. Ovadyahu, and M. Pollak, Phys. Rev. B \textbf{48},
  15025 (1993).
  
 \bibitem{gdcanya97}
  G. Martinez-Arizala, D. E. Grupp, C. Christiansen, A. Mack, N.
  Markovic, Y. Seguchi, and A. M. Goldman, Phys. Rev. Lett. \textbf{78},
  1130 (1997)  
  
 \bibitem{gcdnaa98}
   G. Martinez-Arizala, C. Christiansen, D. E. Grupp,
  N. Markovic, A. Mack, and A. M. Goldman, Phys. Rev. B \textbf{57},
  R670 (1998). 
  
 \bibitem{zm97}
  Z. Ovadyahu and M. Pollak, Phys. Rev. Lett. \textbf{79}, 459 (1997) 
 \bibitem{azm98}
  A. Vaknin, Z. Ovadyahu, and M. Pollak, Phys. Rev. Lett. \textbf{81}, 669
  (1998).
    
 \bibitem{z17}
  Z. Ovadyahu, Phys. Rev. B \textbf{95}, 134203 (2017).
  
 \bibitem{z18}
  Z. Ovadyahu, Phys. Rev. B \textbf{97}, 214201 (2018).
  
 \bibitem{z19}
  Z. Ovadyahu, Phys. Rev. B \textbf{99}, 184201 (2019).      

\bibitem{vz07}
 V. Orlyanchik and Z. Ovadyahu, Phys. Rev. B \textbf{75}, 174205 (2007).
 
\bibitem{jtcvll20}
 J. Delahaye, T. Grenet, C. A. Marrache-Kikuchi, V. Humbert, L. Berg\'{e} and L. Dumoulin, SciPost Phys. 
 \textbf{8}, 056 (2020). 
 
 \bibitem{mz03}
  M. Pollak and Z. Ovadyahu, Phys. Status Solidi C \textbf{2}, 283 (2003).   
  \bibitem{ayy08}
  A. Amir, Y. Oreg and Y. Imry, Phys. Rev. B. \textbf{77} 165207 (2008).    
    
 \bibitem{car94} 
  C. Sagui, A. M. Somoza and R. C. Desai, Phys. Rev. E \textbf{50}, 4865 (1994).
  
 \bibitem{rp96} 
  R. W. Cahn and P. Haasen, Eds., Physical Metallurgy, North-Holland, Amsterdam, (1996).
  
 \bibitem{aa93}
  A. Maheshwari and A. J. Ardell, Phys. Rev. Lett. \textbf{70}, 2305 (1993).   
  
 \bibitem{pw09}
  S. Puri and V.K. Wadhawan (Editors), \textit{Kinetics of Phase Transitions} (CRC Press, Boca Raton, 2009).  
                                                        
\end{thebibliography}
\end{document}